\documentclass{aa}  

\usepackage[T1]{fontenc}
\usepackage{ae,aecompl}

\usepackage{graphicx}
\usepackage{txfonts}
\usepackage{amsmath}
\usepackage{amssymb}	% Extra maths symbols
\usepackage{mathpazo}

\usepackage{float}
\usepackage{psfrag}
\usepackage{ctable}
\usepackage{enumerate}
\usepackage{url}
\usepackage{algorithm}
\usepackage{algpseudocode}
\usepackage{comment}

\newcommand{\plk}{{\it Planck~}}

\usepackage{natbib,twoopt}
\usepackage[breaklinks=true]{hyperref} %% to avoid \citeads line fills
\usepackage{cleveref}

\bibpunct{(}{)}{;}{a}{}{,}             %% natbib format for A&A and ApJ
\makeatletter
  \newcommandtwoopt{\citeads}[3][][]{\href{http://adsabs.harvard.edu/abs/#3}%
    {\def\hyper@linkstart##1##2{}%
     \let\hyper@linkend\@empty\citealp[#1][#2]{#3}}}
  \newcommandtwoopt{\citepads}[3][][]{\href{http://adsabs.harvard.edu/abs/#3}%
    {\def\hyper@linkstart##1##2{}%
     \let\hyper@linkend\@empty\citep[#1][#2]{#3}}}
  \newcommandtwoopt{\citetads}[3][][]{\href{http://adsabs.harvard.edu/abs/#3}%
    {\def\hyper@linkstart##1##2{}%
     \let\hyper@linkend\@empty\citet[#1][#2]{#3}}}
  \newcommandtwoopt{\citeyearads}[3][][]%
    {\href{http://adsabs.harvard.edu/abs/#3}
    {\def\hyper@linkstart##1##2{}%
     \let\hyper@linkend\@empty\citeyear[#1][#2]{#3}}}
\makeatother
\begin{document}

   \title{WHIM-hunting through cross-correlation of optical and SZ effect data in the Virgo cluster filaments}
   \titlerunning{Virgo filaments}

   \author{Cagri Erciyes
          \inst{1}\fnmsep\thanks{E-mail: cagri.erciyes@dlr.de}
          , Kaustuv Basu\inst{1},
          Suk Kim\inst{2}, Soo-Chang Rey\inst{2}
          }

   \authorrunning{C. Erciyes et al.}
   
   \institute{Argelander Institut f\"ur Astronomie, Universit\"at Bonn, Auf dem H\"ugel 71, 53121 Bonn, Germany 
         \and
             Department of Astronomy and Space Science, Chungnam National University, 99 Daehak-ro, Daejeon 305-764, Korea
             }

   \date{Received YYY; Accepted XXX; in original form ZZZ}

% \abstract{}{}{}{}{} 
% 5 {} token are mandatory
 
  \abstract
   {The physical state of most of the baryonic matter in the local universe is unknown, which is commonly referred to as the ``missing baryon problem". It is theorized that at least half of these missing baryons are in a warm-hot, low-density phase outside of the virialized dark-matter halos.}
   {We make an attempt to find the signature of this warm-hot intergalactic medium (WHIM) phase in the filaments of the nearby Virgo cluster by using optical and Sunyaev-Zeldovich effect data.}
   {Specifically, we use a filament-galaxy catalog created from the HyperLeda database and an all-sky Compton-$y$ map extracted from the {\it Planck} satellite data for 2-dimensional cross-correlation analysis by applying spherical harmonics transform. Significance test is based on the null-test simulations which exploits advanced cut-sky analysis tools for a proper map reconstruction. To place upper limits on the WHIM density in the Virgo filaments, realistic baryonic density modelling within the cosmic filaments is done based on state-of-the-art hydro-simulations, and it's done within the signal-boosting routine.}
   {The cross-correlation signal is found to be too dim compared to the noise level in the {\it Planck} $y$-map. At $3\,\sigma$ confidence level, the upper limit on volume-average WHIM density turns out to be $\langle\,n_e\,\rangle<4\times10^{-4}$ cm$^{-3}$, which is indeed consistent with the WHIM parameter space as predicted from simulations.}
   {}


   \maketitle
%
%-------------------------------------------------------------------

%%%%%%%%%%%%%%%%%%%%%%%%%%%%%%%%%%%%%%%%%%%%%%%%%%

%%%%%%%%%%%%%%%%% BODY OF PAPER %%%%%%%%%%%%%%%%%%

%%%%%%%%%%%%%%%%%%%%%%%%%%%%%%%%%%%%%%%%%%%%%%%%%%%%%%%%%%%%%%%\
\section{Introduction}

The first indication that the universe might be holding $40\%-50\%$ of its baryons in its current epoch within the vast filamentary network of dark matter known as the "cosmic web" came from numerical simulations \citep{CenOstriker2006}. This phase of baryonic matter was given the name WHIM (Warm Hot Intergalactic Material) following its expected temperature and density ranges: $T_e\;\in\;[10^5-10^7]$ K and $n_e\;\in\;[10^{-5}-10^{-8}]$ cm$^{-3}$ (e.g., \citet{Bertone2008, Shull2012, Martizzi2019}), and to distinguish it from the denser and hotter plasma that is bound within the potential well of galaxy clusters. The expected properties of WHIM made it obvious that it would be very difficult to detect it via its direct thermal emission in the soft X-ray bands. Hence UV and X-ray absorption lines were identified as optimal tools for the WHIM search, leading to several detections and upper limits (e.g., \citet{Danforth2008}, \citet{Fresco2020}).  Nevertheless, a full accounting of the local baryons via line absorption method was incomplete, which led to the so called "missing baryon problem".

The onset of wide-area millimeter/submillimeter surveys for detecting the Sunyaev-Zeldovich (SZ) effect in galaxy clusters has provided a new approach to search for the WHIM phase. The SZ effect \citep{sunyaev1970small} is a small spectral distortion on the Cosmic Microwave Background (CMB) caused by scattering off the high-energy thermal electrons embedded within the cosmic structures, and crucially, the linear dependence on density for the SZ signal makes it particularly attractive for the WHIM search (\citet{Mroczkowski2019} for a review). Moreover, SZ and X-ray measurements can be considered a more "direct" probe for the WHIM, than line absorption measurements, as the former two directly constrain the plasma properties without any assumption on its metal content. Taking advantage of the full-sky SZ effect map provided by the \plk satellite \citep{ade2016planck}, a series of papers used the stacking technique to claim detection of the WHIM in the cosmic filaments with roughly $5\sigma$ significance (\citet{Tanimura2019a}, \citet{deGraaf2019}). The stacking technique is a common approach in astronomy where the signal-to-noise ratio of a faint signal is improved by averaging many (often several 1000s) individual sky patches, where the location of the signal is inferred from some other data sets, thereby lowering the constribution from stochastic instrumental noise. In these studies, the location of filaments were inferred from bright, elliptical galaxy locations in the SDSS optical data \citep{Alam2015}. More recently, this stacking technique has been improved and newer filament catalogs based on SDSS data have been used, resulting in an improved detection from the same \plk SZ map \citep{Tanimura2020a}, and also, detections have now been made in the X-ray bands, with wide-area X-ray maps from the ROSAT and eROSITA satellites (\citet{Tanimura2020b}, \citet{Tanimura2022b}).

These above WHIM detection results, based on the stacking technique, provide a statistical understanding of the WHIM properties, as the emissions/SZ imprints from many filaments are averaged into a mean signal. A direct search for WHIM from optically confirmed filaments, around an individual cluster, can also have complementary scientific value, e.g. to understand the specific properties of galaxies within a filament or to infer mass accretion rate from specific filaments onto a clusters. The search of WHIM for individual targets have primarily focused on filaments joining nearby clusters, both in the X-ray (\citet{Eckert2015}, \citet{Reiprich2021}) and in the SZ effect \citep{Planck2013filaments}. But these systems cannot probe the nature of WHIM in filaments in an unbiased way, at best they show the transition from the WHIM phase to the denser intracluster medium (ICM) phase. Similarly, important studies for the transition between the unbound and bound gas around galaxy clusters have been made with detailed profile analysis in the cluster outskirts. Examples for this category of work, using the SZ signal, include an analysis of the Virgo cluster outskirts \citep{Planck2016virgo} as well as stacking a sample of several hundred clusters \citep{Anbajagane2022} from South Pole Telescope data.

In this paper we attempt to measure the SZ signal contributed by WHIM directly within a known and well-studied system: the filaments of the Virgo cluster. As our nearest massive galaxy clusters, the galaxy distribution along the various filaments connecting to the Virgo cluster have been very well studied \citep{Tully1982}. Our tool is direct cross-correlation study in the maps-space between the \plk SZ effect map and the optical galaxy distribution in the Virgo filaments, as has been described in a more recent study using a larger data set of HyperLeda database \citep{Kim2016}. Regarding the paper layout, in \cref{Data products} we describe the two data products, and in \cref{Cross-correlation Analysis} the details of the cross-correlation analysis. We simulate the WHIM signal to predict the expected nature of the cross correlation and to make null-tests for error analysis, which are described in \cref{Simulations}. In \cref{Results} the actual results based on \plk data are presented, and some discussion $\&$ summary points with possibilities for future improvement outlined in \cref{discussion_and_conclusions}.

We assume a standard flat $\Lambda$CDM cosmology (Planck2013 results in \citet{ade2014planck}, $\Omega_m$=0.31, $H_0$=67.8 km s$^{-1}$ Mpc$^{-1}$) to convert the redshifts listed in the optical galaxy catalog to distances. We discuss in \cref{discussion_and_conclusions} if the resulting choice of the Virgo-centric distance could have any influence on our correlation analysis. 
%distances without any model correction including conversion of heliocentric radial velocities to the centroid of the Local Group (LG), and also infall of the LG towards Virgo as we don't have accurate kinematics model of all Virgo filament galaxy members (we use a mean Virgo-centric distance of $\approx$18 Mpc via direct linear dependence between redshift and distance). Peculiar velocities of the Virgo filament-hosted galaxies in the vicinity of a massive attractor become significant, and might even dominate this kind of model correction which will become quite uncertain \citep{kashibadze2020structure, castignani2022virgo}. Indeed, we expect a minimal impact on our cross-correlation analysis as the modelled $y$-map should stay approximately the same when there is a slight line-of-sight distance changes for the galaxies. Such effects are already compensated with our null-test algorithm (\cref{null_tests}) which is robust with local geometric deviations and also inclusive error analysis (\cref{error_estimation}) which already accounts for such minor projection issues that mimicks smoothing effect.

%%%%%%%%%%%%%%%%%%%%%%%%%%%%%%%%%%%%%%%%%%%%%%%%%%%%%%%%%%%%%%%\
\section{Data products}
\label{Data products}

\subsection{Optical catalog of filament galaxies} \label{filament_galaxies}

%Following \citet{Kim2016} here. (Fig. 1 $\&$ Table 1 should be presented within this part!)

\begin{figure*}
  \center
  \includegraphics[width=0.85\textwidth, trim=0 240 0 240, clip]{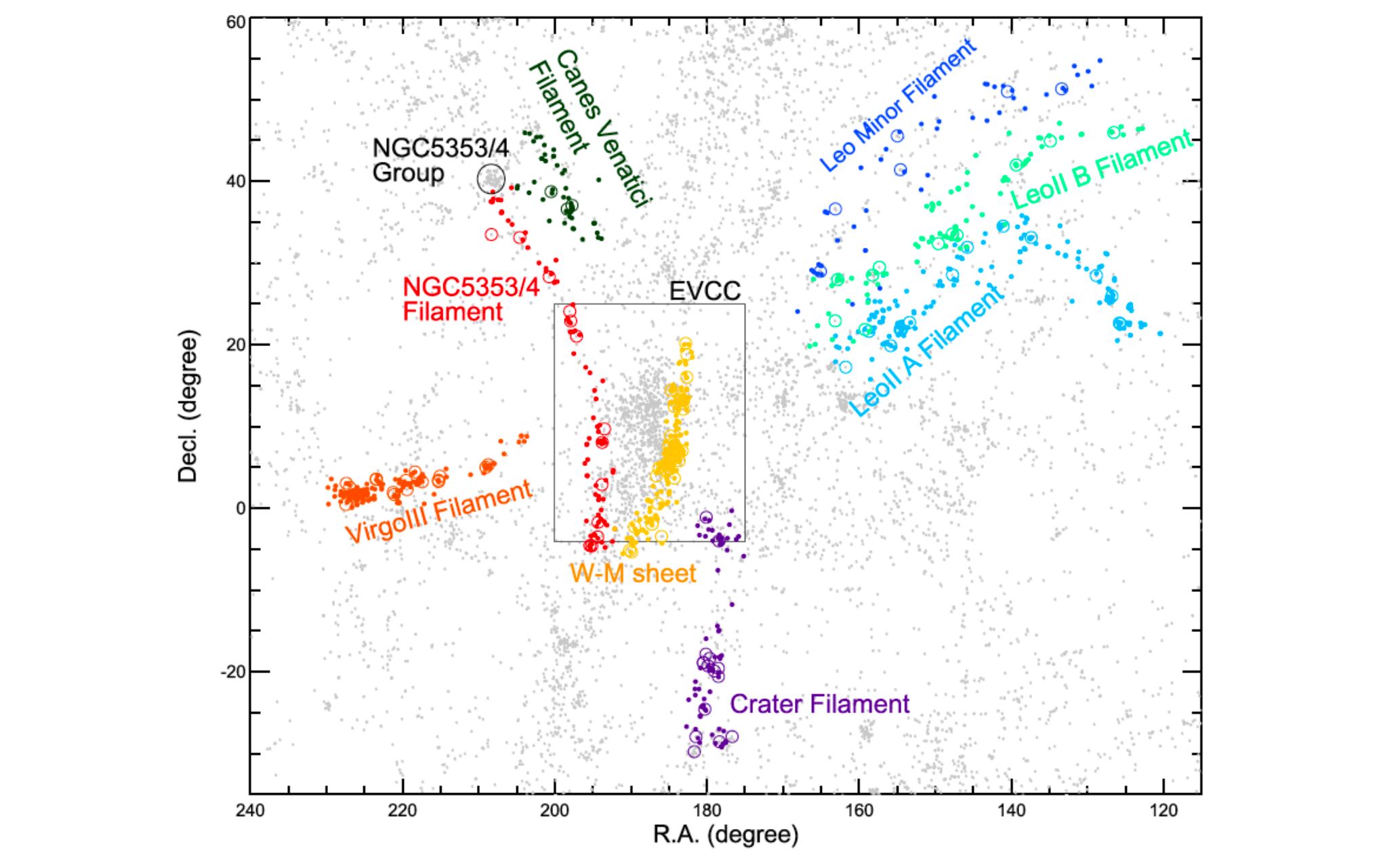}
  \caption{
% Revised by Soo-Chang Rey
Galaxy distribution in the seven filaments and one sheet around the Virgo cluster taken from \citet{Kim2016}. Different colors represents different structures. Bright ($M_{B}$ $<$ $-$19 ) and faint ($M_{B}$ $>$ $-$19 ) galaxies are denoted by large open circles and small filled circles, respectively. The large rectangular box is the region of the Virgo cluster enclosed by the EVCC \citep{Kim2014}.
%    Galaxy distribution in the seven filaments and one sheet around the Virgo cluster. Taken from 
% \citet{Kim2016}
    }
  \label{figure:optical}
\end{figure*}

% Revised by Soo-Chang Rey : 2022, Sep. 24
The filament-like structures (i.e., prolate and oblate overdensities of galaxies) around the Virgo cluster, which is the nearest rich and dynamically young cluster of galaxies \citep{Aguerri2005}, were first studied by \citet{Tully1982}. The most recent and comprehensive attempt to map the large-scale structures around the Virgo cluster exploiting the HyperLeda database \citep{Pature2003} had been carried out by \citet{Kim2016}, whose filament-galaxy catalog is used in this work. A visual representation of these various filaments is shown in Fig.\ref{figure:optical}.

The catalog of galaxies contain 1100 members, grouped into seven filaments and one sheet, as depicted in Fig.\ref{figure:optical}. The search volume and radial velocity cut-offs are discussed in \citet{Kim2016}, to which we refer the readers for further details. One important criterion is the exclusion of the Virgo cluster members, which were defined by the Extended Virgo Cluster Catalog (EVCC; \citet{Kim2014}, see the large rectangular box in Fig.\ref{figure:optical}) and excluded in the current study. A similar region is also masked in the Sunyaev-Zeldovich effect map (see \cref{masking} for details) whose properties are discussed below.  

%The filaments around the Virgo cluster, which is the nearest and rich cluster of galaxies (significantly more massive than smaller galaxy groups like Leo), were first studied by %\citet{Tully1982}. The most recent and comprehensive attempt to map the large-scale structures %around Virgo using Sloan Digital Sky Survey (SDSS DR7, \citet{Abazajian2009}) had been carried out %by \citet{Kim2016}, whose filament-galaxy catalog is used in this work. A visual representation of %the various filament is shown in Fig.\ref{figure:optical}.
%
%The catalog of galaxies contain 1100 members, grouped into seven filaments and one sheet, as %depicted in Fig.\ref{figure:optical}. The search volume and radial velocity cut-offs are discussed %in detail in \citet{Kim2016}, to which we refer the readers for further details. One important %criterion is the exclusion of the Virgo cluster members, which were defined in the Extended Virgo %Cluster Catalog in \citet{Kim2016} (marked by the EVCC box in Fig.\ref{figure:optical}) and excluded %in the current study. A similar region is also masked in the Sunyaev-Zeldovich effect map whose %properties are discussed below.  

\subsection{Planck all-sky $y$-map}

%\plk 2015 NILC $y$-map. 
The very large extension of the Virgo cluster and its filaments on the sky means that only a space-based millimeter/submillimeter telescope can recover the full extended signal of the thermal Sunyaev-Zeldovich (tSZ) effect, and currently, the best data comes from the all-sky survey of the \plk satellite \citep{ade2011planck}. In addition to the sky coverage, the multiple frequencies of the \plk mission enables optimal separation of foregrounds, leading to a cleaner reconstruction of the tSZ signal. We use the all-sky Compton-$y$-parameter maps \citep{aghanim2016planck-tSZ}, where the $y$-parameter is the frequency-independent amplitude of the tSZ signal and is defined as the line-of-sight integral of the electron pressure \citep{sunyaev1970small}
\begin{equation} \label{eq:compton_y}
y = \int n_{e}\frac{k_{B}T_{e}}{m_{e}c^2}\sigma_{T}dl
\end{equation} 
where $\sigma_{T}$ is the Thomson cross-section, $n_{e}$ is the electron number density, $T_{e}$ is the electron temperature, $k_{B}$ is the Boltzmann constant and $m_{e}c^{2}$ is the electron rest mass energy. A visual representation of the $y$-map, on the half sky-sphere centered on the Virgo cluster, is shown in the left image of Fig.\ref{figure:ymaps_original}.

The extraction of the $y$-signal from multifrequency data of \plk is done via separate enhanced modifications of the Internal Linear Combination (ILC) method, which assumes individual maps as linear superposition of multiple sky components which are spatially uncorrelated (see \citet{collaboration2013planck} for an overview). The two main methods are knows as MILCA \citep{hurier2013milca} and NILC \citep{delabrouille2009full}. We specifically use the publicly available \plk NILC maps for our analysis, since its lowest-multipole window function provide better response to the largest-scale emission (see Fig.1 in \citet{aghanim2016planck-tSZ}). However, we also compared our results with MILCA maps and obtain similar upper limits, which is to be expected since both these maps have similar noise properties, leading to similar constraints in the case of non-detections. 

%%%%%%%%%%%%%%%%%%%%%%%%%%%%%%%%%%%%%%%%%%%%%%%%%%%%%%%%%%%%%%%
\section{Cross-correlation Analysis}
\label{Cross-correlation Analysis}
\subsection{Spherical Harmonics}

Correlation measure is one of the most popular techniques (beside stacking approach) to reveal hidden information for the case when the targeted signal is below the system noise level. Correlation of fields measures the expectation value of the products of these fields at different spatial points, in other words provides the strength of link between physical quantities in specified scale. In the case of two-point correlation function, structures can be quantified by the correlator using the expectation value integration
\begin{equation}
\xi(x) \equiv \langle f\left(\Vec{y}\right)f\left(\Vec{x}+\Vec{y}\right)\rangle 
\end{equation}
where the average is taken over all positions $\vec{y}$ and all orientations of the separation vector $\vec{x}$ on field function $f$. For that definition, two-point correlation function would be only dependent on the scale length $\left| \Vec{x} \right|$ by taking advantage of spherical symmetry feature of the field as a manifestation of homogeneity and isotropy characteristics on large scales. Besides, if the field density preserves the Gaussianity, whole statistical information of the corresponding probability density function is contained in $\xi(x)$.  %so there is no need to use any clustering statistics beyond the two-point statistics due to Wick's theorem \citep{bernardeau2002large}. 

%Also, as power spectrum is given by the average of square of the Fourier amplitudes over the field, one can calculate correlation function by taking Fourier transform of corresponding power spectrum. Indeed, as stated in \citet{schneider2014extragalactic}

The correlation function can be calculated by taking the Fourier transform of the corresponding power spectrum. However,
the Fourier modes are not appropriate for describing functions defined over the sphere, at least not for large scales. Instead, one needs a complete set of orthonormal basis on the sphere which is provided by the spherical harmonics. That is the reason why we used \textcolor{gray}{PYTHON} module \textcolor{gray}{HEALPY\footnote{\url{https://healpy.readthedocs.io/en/latest/}}}, which is based on HEALPix \citep{gorski1999healpix},  to be able to properly manipulate pixelated data on the sphere and apply spherical harmonics transforms. Besides this software, we also make use of other cut sky analysis tools such as \textcolor{gray}{PyMaster\footnote{\url{https://namaster.readthedocs.io/en/latest/}}} and \textcolor{gray}{PolSpice\footnote{\url{http://www2.iap.fr/users/hivon/software/PolSpice/}}} which have additional features for generating full sky power spectrum from pseudo power spectrum. In particular, \textcolor{gray}{PyMaster} \citep{hivon2001master} to implement statistical null tests which is described in \cref{null_tests}.
\par

We compute the angular cross-correlation function (CCF) from the cross-correlation power spectrum using the standard definition as follows (see Appendix \ref{app:corr_function} for details):
\begin{equation} \label{correlation_func}
\xi\left(\Vec{x}\right) \equiv \langle f\left(n\right)f\left(n'\right)\rangle = \sum \limits_{l} \frac{2l+1}{4\pi} C_l P_{l}\left(\cos(\theta)\right)
\end{equation}
For our case, where we analyze the cross-correlation signal between two maps, then the Fourier amplitudes in Eq. \ref{a_lm} can be calculated for both of these maps and used for the estimator of angular cross-power spectrum
\begin{equation} \label{Cl_cross}
\hat{C}^{yg}_{l} = \frac{1}{2l+1} \sum \limits_{m} \hat{y}_{lm} \hat{g}^*_{lm}
\end{equation}
where $y_{lm}$ and $g_{lm}$ are the Fourier amplitudes which are calculated from the NILC y-map and galaxy-density map, respectively. This estimator doesn't handle the impact of a pixel masking properly (masked pixels are directly considered as zero-valued). If conversion from the power spectrum of the pixelated map $C^{pix}_l$ to the hypothetical unpixelated one $C^\mathrm{unpix}_l$ is needed then one should use effective pixel window function (see Appendix B in \citet{gorski1999healpix} $\&$ Eq. 15 in \citet{hivon2001master}). 

Effect of masking should be taken into account and this will lower the signal power (e.g., see Fig. D.2 in \citet{aghanim2016planck}). The correction calculation roughly can be done by introducing effective sky fraction, $f_\mathrm{sky}$, which is included as a multiplicative factor in Eq. \ref{Cl_cross} as $f^{-1}_{sky}$ \citep{hill2014detection}. However, $f_{sky}$ correction is only true at the first order and applying a cut sky couples the harmonic modes between them. That is the reason why mode-mode coupling kernel is introduced and corrected using \textcolor{gray}{PyMaster} to obtain an unbiased estimate of the full sky power spectrum (Eq. 14 $\&$ 15 in \citet{hivon2001master}) of a given masked field. The details of the masking procedure is outlined in Section \ref{masking}. Ignoring this fact might induce bias in the derived parameters from likelihoods on the estimated $\hat{C_{l}}$.  \par

For checking the statistical significance of the cross-correlation signal coming from the specified sky patch around the Virgo cluster by implementing statistical null-tests (\cref{null_tests}), we will only consider comparison of the observed signal with random correlations arise from the same configuration space: same pixelization, masking $\&$ two-point statistics . Smoothing effect (instrumental beam and finite pixel size) corrections doesn't need to be applied here and then we exploit \textcolor{gray}{PyMaster} package (\cref{null_tests}) for generating random realizations.

\begin{figure*}
  \includegraphics[width=1\textwidth]{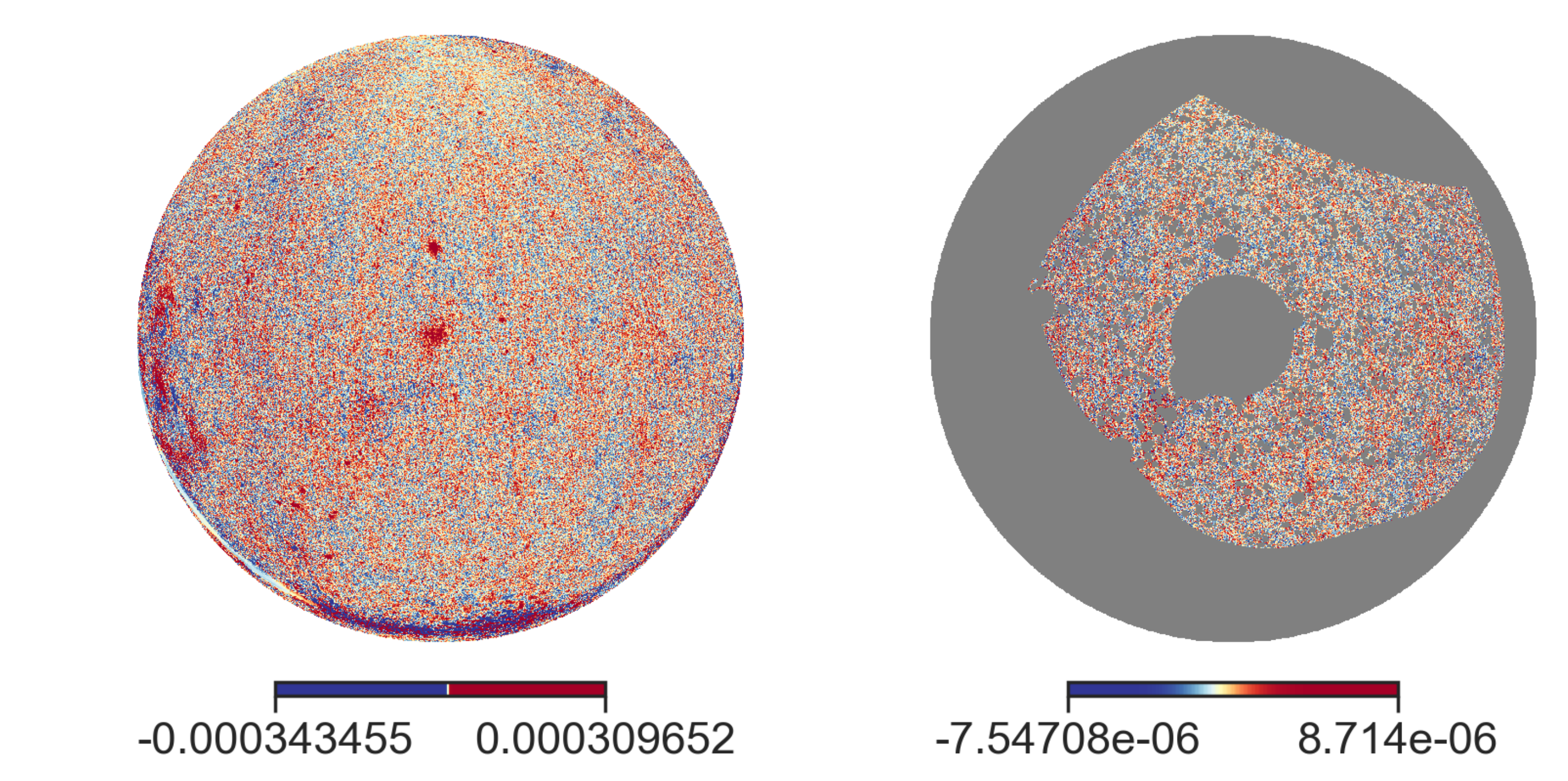}
  \caption{\textit{Left}: NILC $y$-map of the half-sphere which is centered on the Virgo Cluster. \textit{Right}: Masked version of the left map (\cref{masking}). The color-scale denotes the value of the Comptonization parameter, as inferred from multifrequency \plk data.}
  \label{figure:ymaps_original}
\end{figure*}

\subsection{Masking procedure} \label{masking}

In our work we aim to probe low-density regions of the cosmic web, i.e., diffuse WHIM gas located within the filaments around Virgo cluster. We thus mask other possible sources of correlation as much as possible. First, to localize our study, we remove the region outside of the filaments' sky patch which correspond to RA $\approx [120, 230]$ $\&$ DECL $\approx [-30, 55]$ range of degrees. Then, besides the sky patch cropping, 60$\%$ galactic plane mask based on 353 GHz emission after CMB subtraction with no apodization (foreground galactic emission masks)\footnote{\label{foreground_masks}\url{https://irsa.ipac.caltech.edu/data/Planck/release_2/ancillary-data/}} is applied to avoid residual galactic foreground contribution induced by diffuse thermal dust emission mainly. 

As we are only interested in unbound diffuse gas within the filamentary structures located outside of virialized halos, the galaxy clusters detected in the Planck tSZ map (PSZ2, \citet{ade2016planck}), MCXC catalogue \citep{piffaretti2011mcxc} and SDSS DR9 galaxy clusters optical catalogue \citep{banerjee2018optical} are masked out. Galaxy cluster mask is applied with the same strategy as in \citet{tanimura2019detection}, by setting masking radius to three times the cluster radius ($3 x R_{500}$). Specifically for Virgo cluster, to eliminate the impact of dynamical association of the filaments with their parent cluster, e.g. possible reverse shocks that are generated by infalling matter along the Virgo filaments, we apply a more conservative mask whose radius is 3 times the angular size of $R_{200}$ of Virgo cluster. For consistency we adopt the same virial radius $R_{vir}$ as in \citet{urban2011x}, corresponds to $\sim$1.08 Mpc. with a projected radius of $\sim{3^{\circ}.9}$ in the sky. For Planck SZ clusters without assigned radius, mask with a 10 $arcmin$ of radius corresponding to the Planck beam of the NILC $y$-map is exploited.

%At this point, the intention is just eliminating the contamination of highly possible clumped regions of unrelaxed outskirts linked to this dynamically young forming cluster \citep{urban2011x} around where we are trying to probe low-density diffuse WHIM gas within the filaments. In addition, \citet{Kim2016} stated that Virgo Cluster is elongated along a direction parallel to the line-of-sight, owing to the preferential infall of matter through the cosmic web.  

In addition to the contamination from clusters, there might exist a residual emission due to the point sources. In the reconstructed tSZ maps, radio sources appear as negative peaks while infrared sources show up as positive peaks mimicking the signal from clusters. To avoid contamination from these sources, the union of individual frequency point source masks (HFI $\&$ LFI point source foreground masks)\textsuperscript{\ref{foreground_masks}} is used as suggested in \citet{aghanim2016planck-tSZ}. \par

The union of each above-mentioned mask is applied onto both Compton $y$-map and galaxy density map before cross-correlation computation. Mask apodization is redundant for the purpose of spherical harmonics expansion on cut sky in which case there is no gradient calculation on the field, so additional apodization is not applied on our mask (in other words, it is type of \textit{top-hat} mask). Masked pixels are ignored by being interpreted as zeros in \textcolor{gray}{HEALPY} spherical harmonics transform operations. After all of these filtering, total unmasked fraction of the sky is $f_{sky}$ $\approx10$ per cent. In Fig. \ref{figure:ymaps_original}, left panel shows the half sphere of NILC $y$-map which is centered on Virgo Cluster and right panel shows masked version of the same map whose mask is gray-colored. 

%In terms of masking efficiency, one should find a balance between noise dominancy $\&$ contamination reduction by changing the mask size. With the purpose of proper statistics in cross correlation analysis, it shouldn't be too large to be able to use enough number of pixels. That is the reason why cluster mask size is fixed to $3 x R_{500}$ as stated in \citet{tanimura2019detection}, and also fine-tuned richness cut($>=40$) is exploited in the SDSS DR9 galaxy clusters optical catalogue. Otherwise, it's killing the correlation statistics by leaving too few pixels. Besides, point source masks haven't been enlarged to reduce effect of relatively few number of strong radio sources in comparison to total unmasked pixel number.

\section{Simulations} \label{Simulations}
\label{expected_CCF}

%Before analysing the outcome of null tests for significancy check of the measured cross-correlation signal between NILC Compton-$y$ map and galaxy density map, it should be better at first to see what kind of signal would be arised from two maps where the cut sky only consists of these Virgo filaments. 
Before analyzing the real data to check the existence of a cross-correlation signal between the NILC $y$-map and the galaxy-density map, it is worth exploring what kind of signal one should expect if there were no noise and the sky signals (in both Compton-$y$ and optical) would only consist of the Virgo filaments. In this $clear$ representation of the cross-correlation signal, there will be no contamination from any other additional source and no random correlation between structures, and it can be used for comparison with actual or boosted (\cref{boosting}) signal and interpreting the effect of noise. 

The galaxy density map is $clear$ by its construction, so pixels outside the smoothed points are set to zero which is equivalent to masking these regions. However, this is not the case for the $y$-map, which contains thermal SZ contribution from all the structures in the universe as well as residuals from other contaminating signals, plus noise. Thus, to get a 
 $clear$ representation of the tSZ signal coming from diffuse gas within the Virgo filaments, we constructed a noise-free filament $y$-map in which the filament topology is directly reflected by the galaxy density map, based on the assumption that galaxies within each filament are distributed around the main spine axis of these structures. A uniform cross-section curved cylinder around each axis is then filled with WHIM for our study. The details of this modelling is explained below. \par

\subsection{Filament Compton $y$-map reconstruction} \label{filament_ymap}

\subsubsection{Sampling the filament gas particles around spine-axes} \label{filament_ymap_1}

\begin{figure} 
  %\center
  \includegraphics[width=0.5\textwidth]{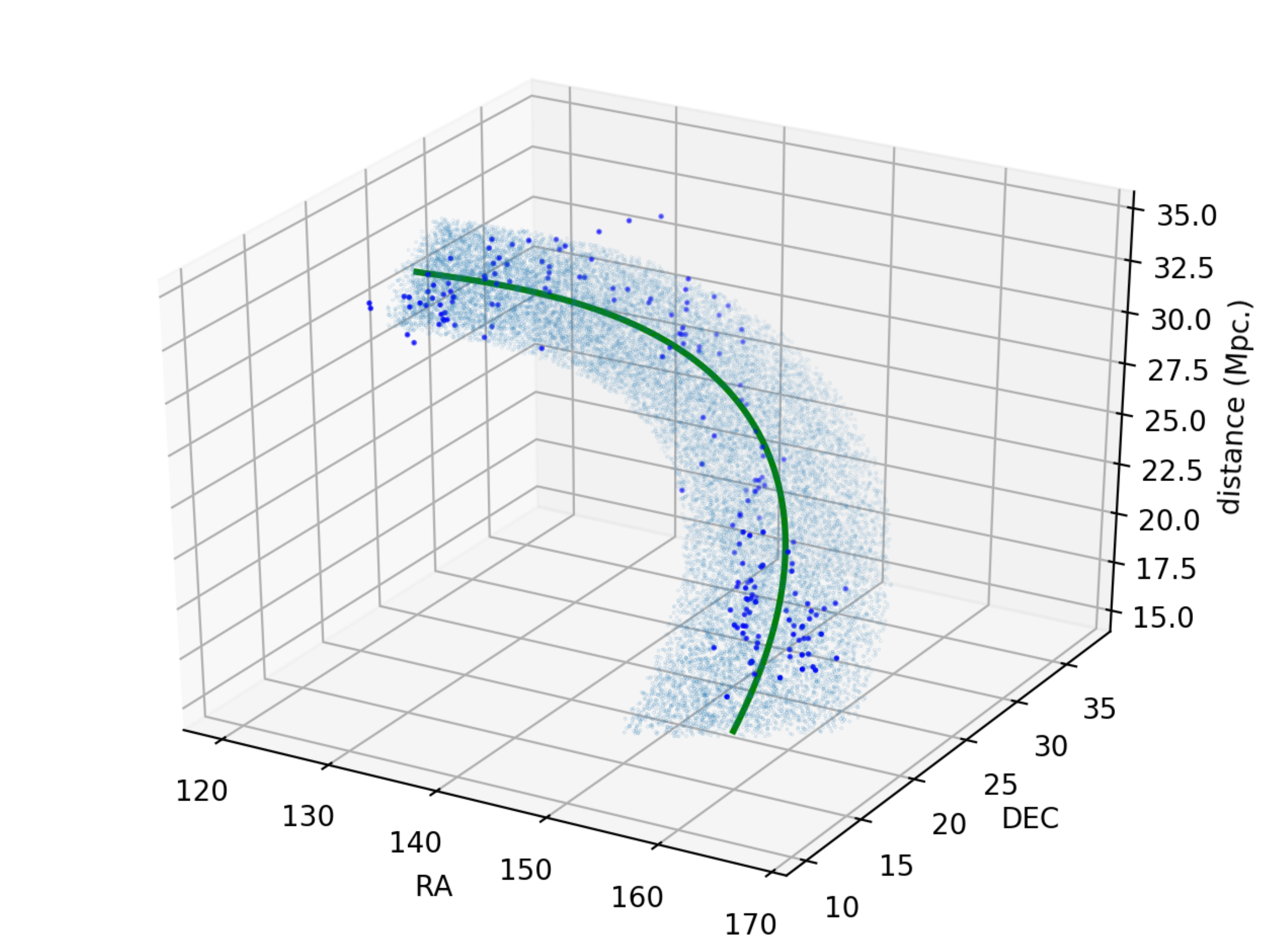}
  \caption{LeoII-A filament in 3-axis: right ascension (RA, in degrees), declination (DEC, in degrees), and line-of-sight distance (in megaparsecs). Filament-hosted galaxies and uniformly sampled gas particles (to simulate WHIM diffuse gas) are shown with dark and light blue points, respectively. The 3-dimensional fitted spine-axis is shown with a green curve around which the gas particles are sampled. Distances of filament galaxies are derived from their heliocentric radial velocities (retrieved from HyperLEDA in \citet{Kim2016}) by assuming a linear relationship between redshift and distance by applying a standard flat $\Lambda$CDM cosmology (Planck2013 results in \citet{ade2014planck}, $\Omega_m$=0.307, $H_0$=67.8 km s$^{-1}$ Mpc$^{-1}$). }
  \label{figure:filament_sampling}
\end{figure}

For simulating the pressure profiles for the diffuse WHIM within known filaments, the first operation would be to sample the gas particles around the main spine-axis of these filaments. The toy model is used in a way that single filament volume-shape is considered as a regular curved cylinder. The algorithm consists of two parts: three-dimensional fit to extract main spine axis of filament and uniform sampling around the main spine-axis. Each of these steps are outlined below. 

For three-dimensional fit to positions of galaxies at each filament, least square method with the option of Huber Loss \citep{huber1992robust} is used, to minimize outliers' influence in the fitting process. This robust loss function is widely used in the context of machine learning applications, among others. Then, cubic spline interpolation is applied onto fitted points and smooth three-dimensional fit for the main spine axis is obtained. For each filament, spine axis consists of N ordered points, i.e. $x_{1}, x_{2}, ...x_{N}$. To be able to perform suitable rotation and scaling operations, a Cartesian coordinate transformation is applied onto the axis points. Sampling is iteratively done for each point along a spine-axis, and these points act as a center of constructed discs. In each iteration, particles are uniformly sampled around these centers in the xy plane, and then uniform random perturbation along the z axis is added onto the sampled points. In this way, uniform disc whose thickness equal to $|x_{n+1} - x_{n}|$ and whose normal lies along z axis is constructed. Then, the constructed disc is rotated by using 3-dimensional rotation matrix and linearly transformed to the exact location $x_{n}$ on the spine axis. 
%In that case, disc normal axis direction is matched with the base vector which corresponds to spine gradient on $x_{n}$ as merged discs should follow the main spine axis at the end. 
At each point interval on the spine axis, 30000 particles are used for sampling. At the end, we have roughly 40 million particles for all filaments in total. 

There is one sheet-like structure besides the filamentary structures in our sample, which is called the \textit{W-M sheet}, including galaxies from the W and M clouds of the Virgo cluster \citep{Kim2016}. It is interpreted as a filament too, whose spine axis is chosen to be nearly perpendicular to line-of-sight. The effect of this approximation for the W-M sheet-like structure would be negligible in our cross-correlation analysis, because it lies behind the Virgo cluster and thereby most of the region is masked.\par

In sampling process, radius of the discs, in other words filament radius is determined by using scaling relations from  \citet{gheller2015properties} and \citet{gheller2016evolution}. Since the filament lengths are already known, we can use the scaling relation between enclosed gas mass $\&$ estimated length of filaments, which is shown in Fig. 8 in \citet{gheller2015properties} and Fig. 3 in \citet{gheller2019survey}. 
%we can estimate the gas mass range of the filaments. 
While the filament length increases, the scatter within the relations become much less in above figures. For this reason, the longest filament (NGC 5353/4) whose length is $\approx{22}$ Mpc \citep{Kim2016} chosen to estimate gas mass range which then can be used to determine reasonable filament radii. For $\approx{22}$ Mpc., relevant mass range is roughly $10^{12.7} - 10^{13.7} M_{\odot}$. Then, relation between the enclosed gas mass 
%and volume for the ﬁlaments (Fig. 7 in \citet{gheller2015properties}) or enclosed gas mass 
and radius of the filaments (Fig. 9 in \citet{gheller2015properties}) is used to get an estimated radius range for NGC 5353/4, which is roughly $0.4 - 1.2$ Mpc. The median value within this range, which equals to 0.8 Mpc, is assigned as the filament radius for NGC 5353/4. 
%Actually, we don't need to be so fussy for obtaining accurate $\&$ precise filament radius values at this section because we just try to obtain an expected normalized shape of the correlation function between galaxy density map $\&$ filament $y$-map. Thus, coherently estimated $y$-values should be enough for this kind of analysis here. 
We note, however, that accurate assignment of filament radius is not very important for our study, as we are interested in only the approximate shape and normalization of the cross-correlation function. 
%The uncertainties related to determining filament radii is addressed in \cref{error_estimation}.  
\par

Then, the remaining issue is estimating radius ratios for the filamentary structures and one can directly use the scaling relation between filament gas mass $\&$ volume again. However, this approach may enhance the deviations in relations because of the filaments with short lengths in our sample. Hence, we use another way to get ratios of filament radii following \citet{gheller2016evolution}. 
%which seems also more proper according to the method being used for obtaining filamentary structures. 
In their Fig. 18, there is a tight correlation between host filament total gas mass $\&$ total mass of galaxies, and also between the length of the host filament $\&$ total mass of galaxies. We use these to re-check the mass range we get for NGC 5353/4. According to length of 22 Mpc., related galaxy mass range is roughly $10^{12} - 10^{13} M_{\odot}$ and this corresponds to filament gas mass range $10^{12.7} - 10^{13.7} M_{\odot}$. It is shown that mass of the filaments tends to grow almost linearly with that of the resident galaxy halos within filaments. 
%In addition, Fig. 13 in \citet{gheller2016evolution} also shows correlation between filament backbone length and number of galaxies, fetched from both survey and simulation data. Two dataset show good agreement in filament lengths larger than 10 Mpc with a large scatter, and number of galaxies that are hosted by NGC 5353/4 is already consistent with the relation from this figure. 

By assuming nearly similar galaxy population properties for all filamentary structures in average (it's underlined in \citet{Kim2016} that all filaments mostly consist of faint galaxies with $M_{B} > -19$ which corresponds to $\approx$88\% of the total filament-hosted galaxy sample, and bright galaxies exhibit uneven distribution with sheer amount of gaps along their host filament), we can expect a linear correlation between filament gas mass and number of galaxies hosted by filament. By making benefit of this approach, and considering filament shapes as cylinder, filament gas mass is related with $LR^2$, where $L$ is backbone length and $R$ is filament radius. This way, we obtain radius ratios of these filaments whose spine lengths are already known. The rest of the filament radii are determined according to these ratios and also picked reference radius value for NGC 5353/4. 
In Fig. \ref{figure:filament_sampling}, uniformly sampled gas particles around the fitted spine-axis for LeoII-A filament can be seen.

\subsubsection{Temperature selection for isothermal filaments} \label{filament_ymap_2}

\begin{figure*} 
  \includegraphics[width=1\textwidth]{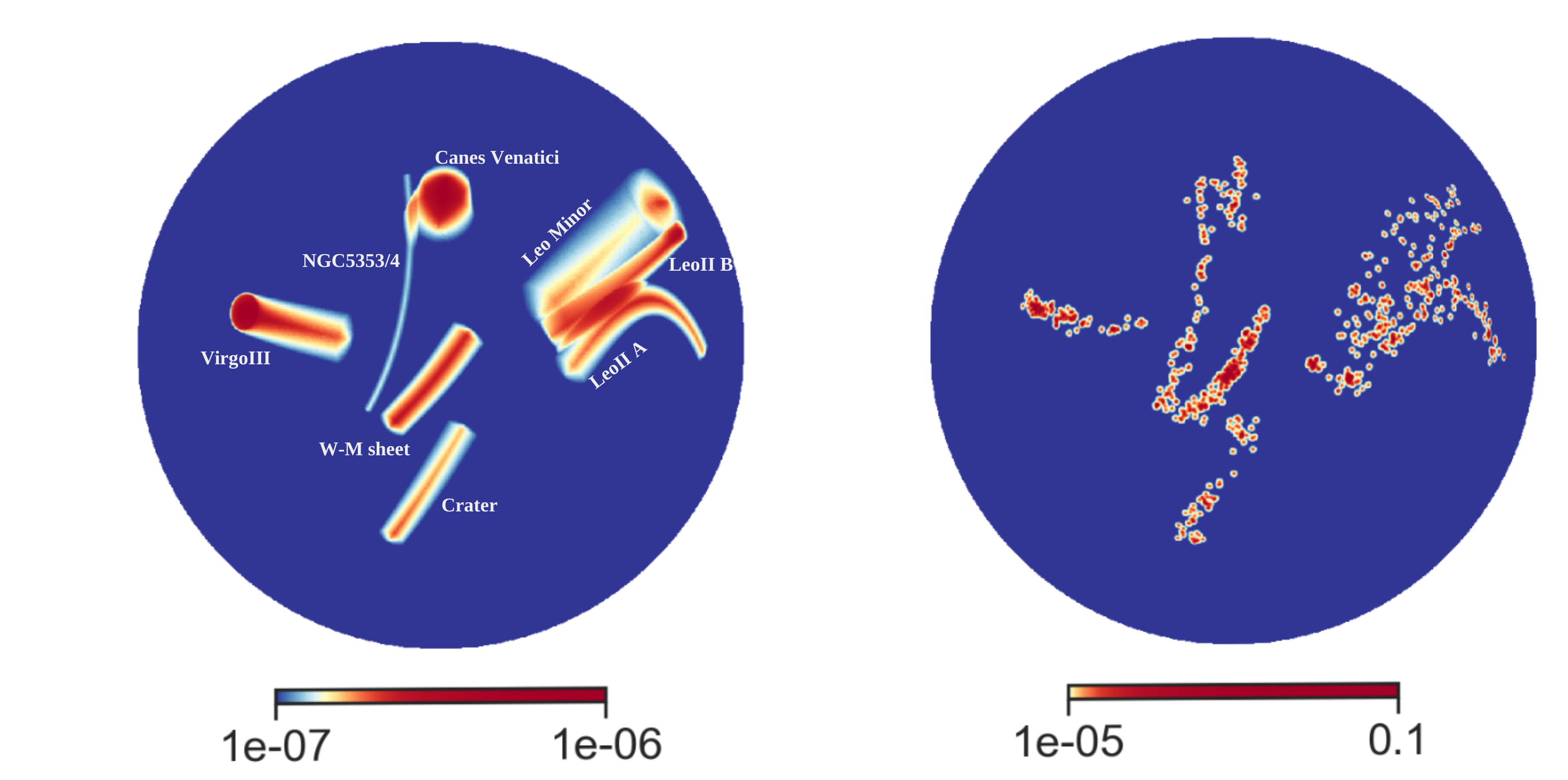}
  \caption{\textit{Left}: Modelled noise-free $y$-map of the filamentary structures around the Virgo Cluster, which is also smoothed with the Planck beam (\cref{filament_ymap}). Projection effects for each stated filament can be realized by comparing with Fig. \ref{figure:optical}. \textit{Right}: Constructed galaxy density map from the point locations (\cref{filament_galaxies}) by using \textit{get$\_$interp$\_$weights()} function in \textcolor{gray}{HEALPY} module and smoothing with a Planck beam whose FWHM is $\approx$30 arcminute. Both maps are centered on the Virgo Cluster.}
  \label{figure:filament_vs_galaxy_map}
\end{figure*}

After obtaining uniformly sampled particles, line-of-sight integral of the pressure needs to be calculated to extract Compton-$y$ value within each pixel on our \textcolor{gray}{HEALPix} sky maps. Firstly, isothermal filament model is assumed, which is also motivated from Fig. 8 in \citet{gheller2019survey} and Fig. 17 in \citet{gheller2015properties} that shows transverse profile of gas temperature as a function of the scale height from the spine axis for the representative filaments. 
%extracted from simulated models which have different adopted baryonic physics. 
It seems that gas temperature remains nearly uniform from the spine to outer parts for most of the models. \citet{gheller2015properties} explains this phenomenon arguing that inner parts of the filaments are thermalized at an almost constant temperature by shock waves propagating from the centre outwards, but the outer part of the filaments can comprise non-shocked cells that lower the average temperature. Besides, it's observed that filaments follow well-defined scaling relations in the gas temperature - mass plane \citep{gheller2015properties}. Within our sample of dynamically active Virgo filaments, expected mass range is $10^{12.5} - 10^{14} M_{\odot}$ and corresponding average mass-weighted gas temperature range is roughly $10^{6} - 10^{6.8}$K, which is shown in Fig. 3 from \citet{gheller2019survey} and Fig. 10 from \citet{gheller2015properties} with a large scatter even for high-mass filaments. Because of this broad dispersion, we didn't apply scaling relation and used the logarithmic median temperature value within the specified range: $10^{6.4}$K is picked as a diffuse gas temperature for each isothermal filament in our sample. The uncertainty in the Compton $y$-parameter calculated over these model parameters is addressed in \cref{error_estimation}. In the context of normalized cross-correlation function shape, the temperature selection only effects the scaling, so not the signal shape itself. \par

\subsubsection{Density modelling and projection} \label{filament_ymap_3}
The Compton-$y$ value in an isothermal gas is simply proportional to line-of-sight integral of the gas density:
\begin{equation} \label{eqn_density_integral}
\bar{n} = \frac{\int ndl}{\int dl}.
\end{equation}
Numerically, $\bar{n}$ can be easily obtained from uniform point sampling under a given density function $n$, and line-of-sight distance can be calculated from the interval between the nearest and furthest particles within each pixel on the map. Regarding to the density profile, beta-model is chosen as the average transverse profile of gas density for selected representative filaments, as shown from simulated models in Fig. 8 from \citet{gheller2019survey} and  Fig. 17 from \citet{gheller2015properties}. It is clear that for simulations with radiative scenario the central mass density becomes roughly 2 - 3 times higher than average, and profile getting steeper due to the higher compression levels \citep{gheller2015properties}. Commonly, all density profiles exhibit smooth drop by a factor of broadly 2-10 at $\sim$0.3$r_{fil}$ from the spine axis, then slowly declining outwards. Uncertainty on the density profile steepness and its effects on the results are discussed in \cref{error_estimation}. With the purpose of CCF estimation and initial null tests, almost shallow density profile is picked and sampled with a quasi-Monte Carlo integration to obtain an expected density value $\mathbf{E}(n)$ which corresponds to $\bar{n}$. Used isothermal $\beta$-model 
%arisen from empirical King approximation \citep{arnaud2009beta} to represent 
the diffuse WHIM gas density profile is written as
\begin{equation}
n_{e}(r) = n_{e}(0)\left[1+\left(\frac{r}{r_{c}}\right)^{2}\right]^{-3\beta/2}
\end{equation} where $r_{c}$ is the core radius of the distribution and $n_{e}(0)$ is the central electron density value. To be able to construct relatively shallow density profiles with smooth drop at inner region and nearly flat outwards for the diffuse gas, we kept $r_c$ and $\beta$ parameters small, e.g. $r_{c} = 0.06$ $\&$ $\beta = 0.17$ holding the relation of $\langle n_{e} \rangle \simeq 0.42 \, n_{e}(0)$,  where $\langle n_e \rangle$ is the transversal (widthwise) electron mean density which corresponds to $\bar{n}$ in Eq. \ref{eqn_density_integral}.  

\par
The final stage to construct noise-free $y$-map from the uniformly sampled isothermal gas particles is the projection. The pseudocode for the algorithm is shown in Algorithm \ref{ymap_pseudocode}. First, for each of the sampled particles, host pixel indices are found by using \textit{ang2pix()} function in \textcolor{gray}{HEALPY} module, and unique list of pixel indices is obtained. Then, for each unique pixel index, particles locating inside the boundaries of the pixel with given indice are fetched and mean density (by simply averaging the local density values which are attached onto particle positions w.r.t. vertical distance from spine axis) $\&$ line-of-sight distance interval is calculated for the obtained set of particles. As a last step, Compton $y$-parameter is calculated by multiplying these quantities with fixed temperature value and rest of the constants as given by Eq.\ref{eq:compton_y}. 
%shown below \citep{sunyaev1970small}
%\begin{equation} \label{eq:compton_y}
%y = \int n_{e}\frac{k_{B}T_{e}}{m_{e}c^2}\sigma_{T}dl
%\end{equation} where $\sigma_{T}$ is the Thomson cross-section, $n_{e}$ is the electron number density, $T_{e}$ is the electron temperature, $k_{B}$ is the Boltzmann constant, $m_{e}c^{2}$ is the electron rest mass energy, and the integration is along the line of sight. 
This calculation needs to be done separately for each filamentary structures because of the linearly additive behaviour of the Compton $y$-parameter in the case of overlap, e.g. node region of LeoII filaments in Fig. \ref{figure:filament_vs_galaxy_map}. After finishing the main loop to get estimated Compton $y$-parameter values over pixels, generated map is smoothed with the Planck beam whose FWHM is $\sim$10 arcminute. 

\begin{algorithm}
  \caption{Construction of filament-based $y$-map}\label{ymap_pseudocode}
  \begin{algorithmic}[1]
      \State $y\_map\gets initialize()$ \Comment{null $y$-map}
      \State $pixel\,indices\gets ang2pix(spatial\,positions)$
      \State $unique\,indices\gets unique(pixel\,indices)$
      \For{\texttt{each unique\,indices}}
        \For{\texttt{each structures}}
            \State $windowed\,particles\gets fetch(unique\,indice)$
            \State $\bar{n}\gets mean\_density(windowed\,particles)$
            \State $l\gets max\_separation(windowed\,particles)$
            \State $\hat{y}\gets calculate\_Compton\_y(\bar{n}, l, T)$ \Comment{Eq. \ref{eq:compton_y}}
            \State $y\_map[unique\,indice] \mathrel{{+}{=}} \,  \hat{y}$
        \EndFor
      \EndFor
      \State $y\_map\gets smooth(y\_map)$ \Comment{with Planck beam}
  \end{algorithmic}
\end{algorithm}

Filament-based constructed smooth $y$-map is presented on the left side of Fig. \ref{figure:filament_vs_galaxy_map}, with the smoothed galaxy density map on the right side. With the purpose of suppressing the sparsity effect on the optical data in which only $\sim$1000 galaxy positions exist within a wide field, galaxy density map is constructed by using \textit{get$\_$interp$\_$weights()} function in \textcolor{gray}{HEALPY} module which returns the 4 closest pixels on the two rings above and below the specified location and corresponding weights. Thereafter, smoothing with a beam whose FWHM is $\approx$30 arcminute is applied onto the map to break high discreteness level. \par 

Projection effects can be obviously seen for filamentary structures in Fig. \ref{figure:filament_vs_galaxy_map}. Leo Minor seems to be the largest filament because of its closeness, indeed it is one of the smallest filaments in terms of the number of hosted galaxies and also its backbone length. LeoII A $\&$ LeoII B are overlapping on the $y$-map sky, observed as a multi-stem structure. NGC 5353/4 is the longest $\&$ most distant filament in our sample extends out from the NGC 5353/4 group, running tangentially past the Virgo cluster rather than pointing toward it which is seen very thin \citep{Kim2016}. Additionally, Canes Venatici is one of the filaments running from the vicinity of the NGC 5353/4 group to the Virgo cluster and its maximal elongation component lies along a SGY direction which is the axis of the supergalactic coordinate system, that roughly coincides with the line of sight from our Galaxy.
%of the earthling observer for this case \citep{Kim2016}.

\subsubsection{CCF calculation}
\begin{figure} 
  \includegraphics[width=0.5\textwidth]{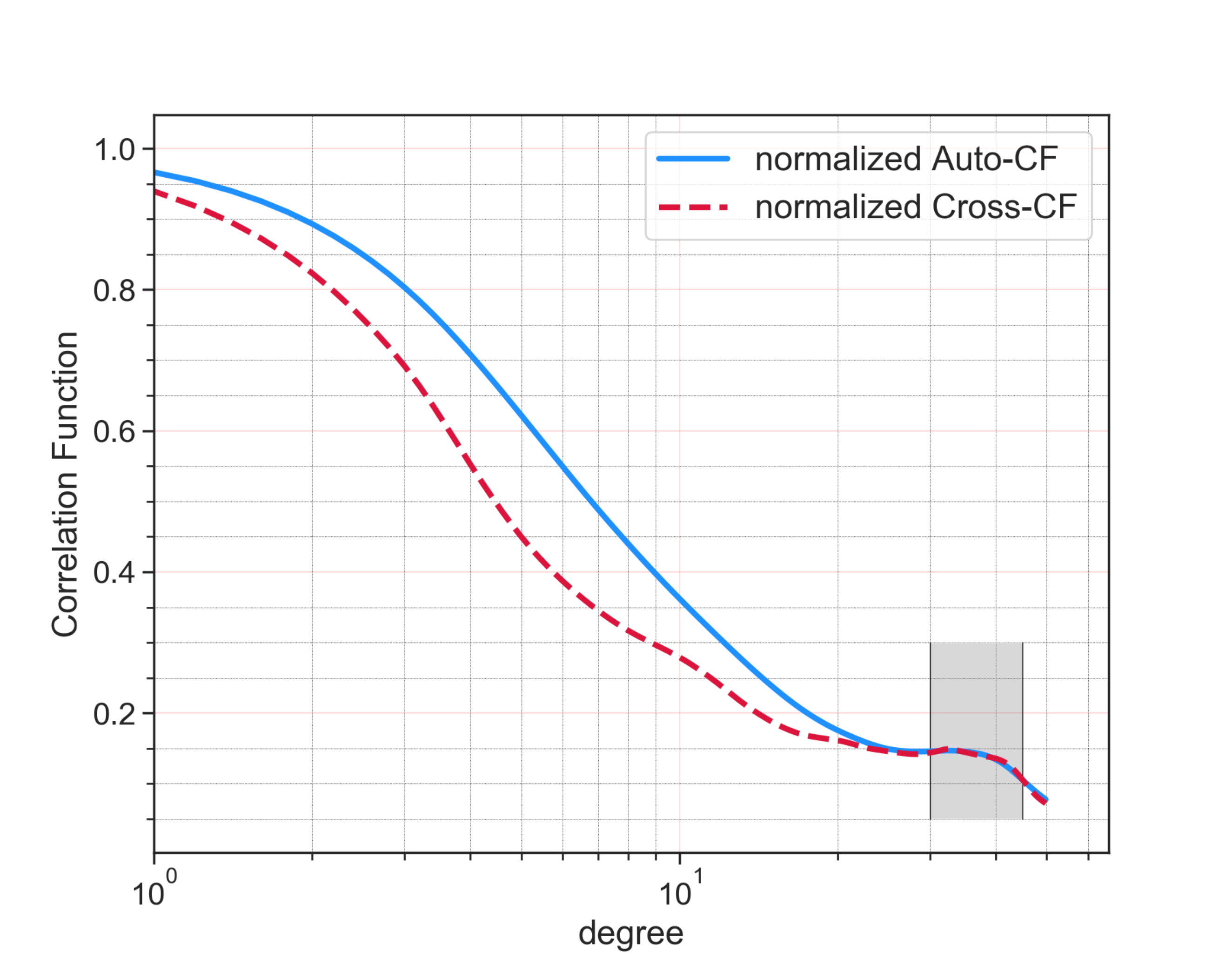}
  \caption{Expected normalized correlation functions for the masked version (\cref{masking}) of the noise-free (and also without any contamination) fields that are shown in Fig. \ref{figure:filament_vs_galaxy_map}. Blue straight line points up the auto-correlation function for the modelled filament $y$-map (left panel in Fig. \ref{figure:filament_vs_galaxy_map}). Red dashed line shows the cross-correlation function between filament $y$-map and galaxy density map (right panel in Fig. \ref{figure:filament_vs_galaxy_map}). The characteristic statistical feature of these filamentary structures in terms of correlation degree is obviously seen as a bump in large scales roughly between 30 and 40 degrees, which is shown with a gray band. This scale corresponds to the average separation between primary elongation occurrences of the filamentary structures, as the bump in correlation function indicates that there should be a specific repetitive interval within the analysed topology.}
  \label{figure:expected_CCF}
\end{figure}

Before retrieving the expected correlation function between these two maps shown in Fig. \ref{figure:filament_vs_galaxy_map}, the cumulative mask, extracted in \cref{masking}, is applied. To calculate \textit{cross}(or \textit{auto})-correlation function, \textcolor{gray}{anafast()} function is used to obtain \textit{cross}(or \textit{auto})-power spectrum, and then \textcolor{gray}{bl2beam()} function exploits this transfer function in spherical harmonic space to compute a circular beam profile in real space, in other words, correlation function. In Fig. \ref{figure:expected_CCF}, normalized cross-correlation function (CCF) between modelled filament $y$-map $\&$ galaxy density map is shown (red dashed line). There is an obvious bump at large scales around 35$^{\circ}$, and to test its validity we also calculated auto-correlation function of the filament $y$-map, which can also be seen in Fig. \ref{figure:expected_CCF} (blue straight line). The exact coincidence of the two bumps confirm its reality, as the characteristic angular scale of the Virgo filaments, which we exploit in \cref{boosting}.

In addition, we perform a null-test for this excess correlation signal at 35$^{\circ}$ scale. Maps consisting of different number of random lines, whose lengths are within the similar range of Virgo filaments, are constructed, and the CCF between these random maps and galaxy density map are calculated. No bump is seen in these null-tests, which confirm our hypothesis as the bump to be a real feature, that we target for our CCF signal identification (\cref{boosting}).
%to check whether the observed bump at large scales in CCF is typical to our field topology or random. There is no any bump observed in this specific \textit{null}-test, 
%so this large-scale bump seems to be a characteristic feature from which we will benefit in \cref{boosting}. 

\subsection{Null-test for random correlations} \label{null_tests}

\begin{figure*} 
  \includegraphics[width=0.96\textwidth, trim=0 100 0 100, clip]{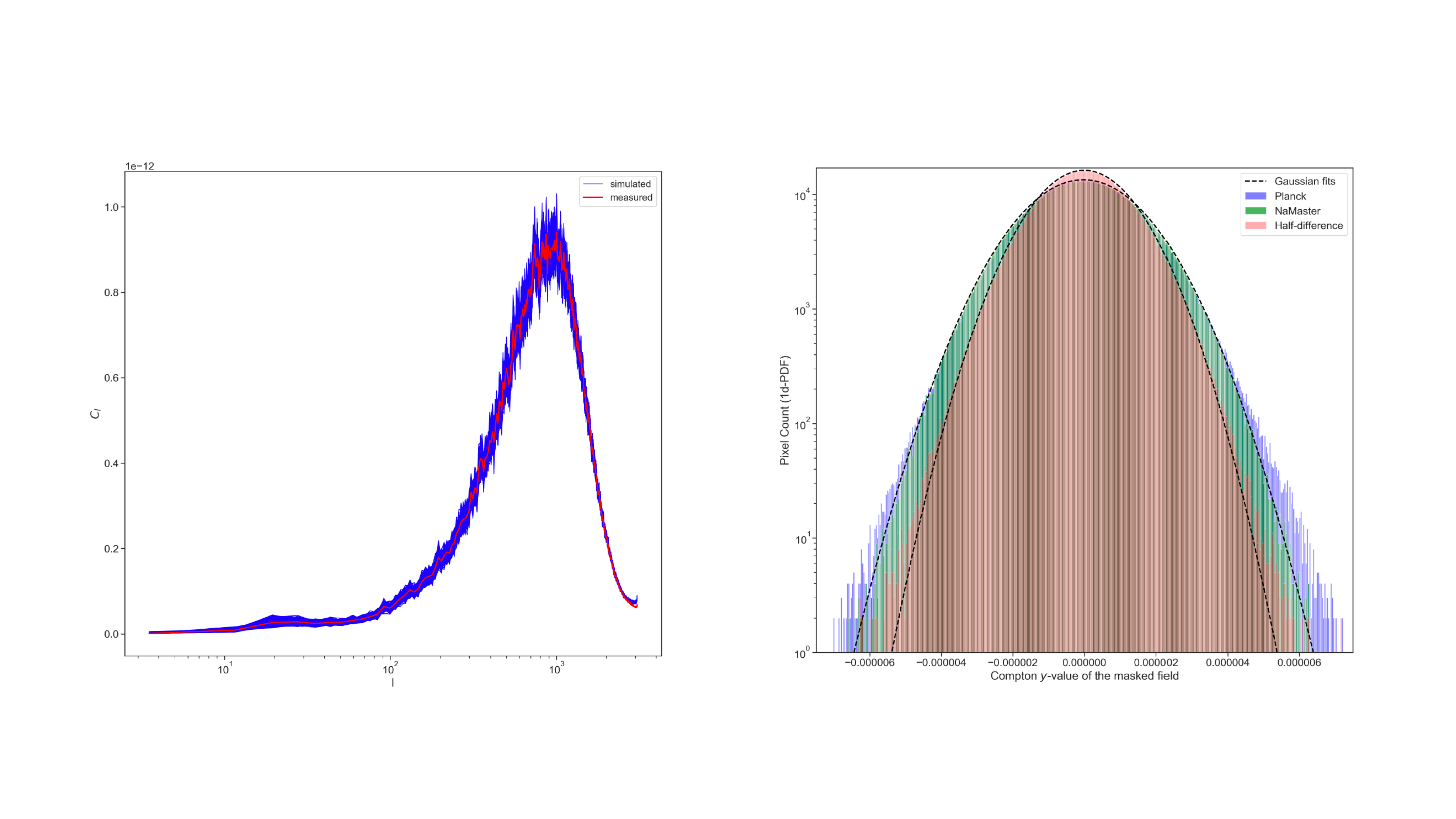}
  \caption{\textit{Left:} Power spectrum comparison between randomly created maps via Algorithm \ref{nulltest_pseudocode} (blue straight lines) and actual NILC $y$-map (red straight line), including the cumulative mask (\cref{masking}). \textit{Right:} 1-dimensional probability density functions (PDF) of the unmasked pixel values for the NILC $y$-map (blue bars), one of the simulated maps (with \textcolor{gray}{NaMaster}) by using the analysed NILC $y$-map field itself (green bars), and one of the simulated maps (with \textcolor{gray}{NaMaster}) by using the half-difference map (red bars). Gaussian fits are sketched for the histograms of randomly simulated maps (black dashed lines). Together with the left panel, it's obvious that usage of the half-difference map, or in other words, approximated white noise is not suitable for the null tests as it's not providing the matching statistics. On the other hand, by taking advantage of \textcolor{gray}{NaMaster}, simulated maps which are obtained by using the own statistics of the analysed field (right panel in Fig. \ref{figure:ymaps_original}) appears to be the best option with the purpose of null test in our case as it mimics the field likewise statistically, which can be directly seen from both 1-d PDF and power spectrum comparisons.}
  \label{figure:Pymaster}
\end{figure*}

With the intent of assessing the significance of the measured CCF signal between NILC $y$-map (right panel in Fig. \ref{figure:ymaps_original}) and smoothed galaxy density map (right panel in Fig. \ref{figure:filament_vs_galaxy_map}), the range of CCF values need to be calculated when there is no real correlation expected between these two maps. For these null tests, we performed Monte Carlo based simulations by creating random realizations of the actual NILC $y$-map by using both \textcolor{gray}{PyMaster} (\textcolor{gray}{NaMaster} python wrapper) and \textcolor{gray}{HEALPY}. The pseudocode is shown in Algorithm \ref{nulltest_pseudocode}. Firstly, to be able to calculate estimated full-sky power spectrum, the fields is constructed by using \textcolor{gray}{NmtField()} function considering NILC $y$-map as the main input field and specifying the mask that is created in \cref{masking} (used in the rest of analysis). Then, full-sky auto-power spectrum of the given masked $y$-map is estimated by using \textcolor{gray}{compute$\_$full$\_$master()} function. This reference $C_{l}$ is used to create 1000 random full-sky maps with \textcolor{gray}{synfast()} function, and the above-mentioned mask is applied onto these simulated maps afterwards. As the last step, CCFs are calculated between these randomly created masked maps and given smoothed galaxy density map. \par

We follow this strategy for the null tests to create random masked maps whose statistics are as much as close to the masked NILC $y$-map in terms of power spectrum and also probability density function (PDF) of unmasked pixels. In Fig. \ref{figure:Pymaster}, on the left panel, power spectrums of the randomly created masked $y$-maps and the one from masked NILC $y$-map are compared. On the right, 1d PDF of unmasked pixels of the NILC $y$-map and one of the randomly created maps via \textcolor{gray}{NaMaster} are presented. In both of the cases, statistics match well enough and it clearly indicates that our approach is well-suited for the null test within the context of CCF significance measure. 

\begin{algorithm}
  \caption{NULL test for CCF significance measure}\label{nulltest_pseudocode}
  \begin{algorithmic}[1]
      \State $random\_maps\gets initialize()$ \Comment{empty list is assigned}
      \State $NULL\_CCFs\gets initialize()$ \Comment{empty list is assigned}
      \State $y\_field\gets construct\_field(NILC\_y\_map,\, mask)$
      \State $full\_sky\_C_{l}\gets compute\_full\_master(y\_field, \, y\_field)$
      \State $simulation\_number\gets 1000$
      \For{\texttt{\textit{i} in range(simulation\_number)}}
        \State $random\_map\gets synfast(full\_sky\_C_{l})$
        \State $masked\_random\_map\gets mask(random\_map, \, mask)$
        \State $random\_maps[i]\gets masked\_random\_map$
        \State $NULL\_CCFs[i]\gets CCF(random\_maps[i], \, galaxy\_map)$
      \EndFor
  \end{algorithmic}
\end{algorithm}

For the purpose of the null test, we also checked the usage of half-difference maps
\begin{equation} 
hd = \frac{y_1 - y_2}{2} = \frac{(S+N_1)-(S+N_2)}{2}
\end{equation} 
where $y_1$ $\&$ $y_2$ are the $y$-maps obtained from NILC-pipeline of the first and second half-ring observations, respectively. $hd$ is the half-difference of the two half-ring $y$-maps, which can be used in noise characterization as it gives an estimate of map noise ($N_1$ $\&$ $N_2$) where astrophysical signal ($S$) is canceled out in the measured data. Additionally, Planck $y$-maps exhibit a highly non-homogeneous noise, so there is an expected strong spatial dependency too. \par

In Fig. \ref{figure:Pymaster}, comparison of the histograms of pixel values between masked NILC $y$-map $\&$ half-difference map is shown on the right panel, with corresponding Gaussian fits to each of them. As it can be seen, 1-dimensional PDF of half-difference map is well described by Gaussian fit as expected with negligible tails in both directions. For the masked NILC $y$-map, fit also well-suited except for the endmost pixel values which indicates the possible contamination as a remaining astrophysical signal from the unmasked structures. The most important outcome from this comparison is that the half-difference noise maps are not really fit to proper null test analyses in our case because statistics is significantly different in terms of 1d PDF. The masked NILC $y$-map has higher variance in comparison with half-difference map due to either NILC-pipeline or residual foreground contamination, and this residual bust be included in the CCF robustness estimation. Hence, direct usage of the measured power spectrum of the masked $y$-map, to create random realizations, provide characterization of both statistical and systematics errors in a more robust way.
\par

To perform the tests, the Algorithm \ref{nulltest_pseudocode} is applied with the primary input of half-difference map as well as the NILC $y$-map to compare range of the CCFs arising from random correlations. Again, 1000 random realizations of the masked half-difference map are generated, and corresponding CCFs with galaxy density map are calculated. In Fig. \ref{figure:results_1}, the cumulative range of null-CCFs are presented for usage of both NILC $y$-map and half-difference map by comparing deviation levels up to 3$\sigma$. In all scales, scattering is much broader for the case with direct NILC $y$-map usage which can be already expected by just looking at 1d PDF results directly in Fig. \ref{figure:Pymaster} as the half-difference map manifests smaller amplitude range and also lower spectral power magnitude.

%%%%%%%%%%%%%%%%%%%%%%%%%%%%%%%%%%%%%%%%%%%%%%%%%%%%%%%%%%%%%%%\
\section{Results} \label{Results}

\subsection{Non-detection with {\it Planck} data}

\begin{figure}[h]
  \includegraphics[width=0.5\textwidth]{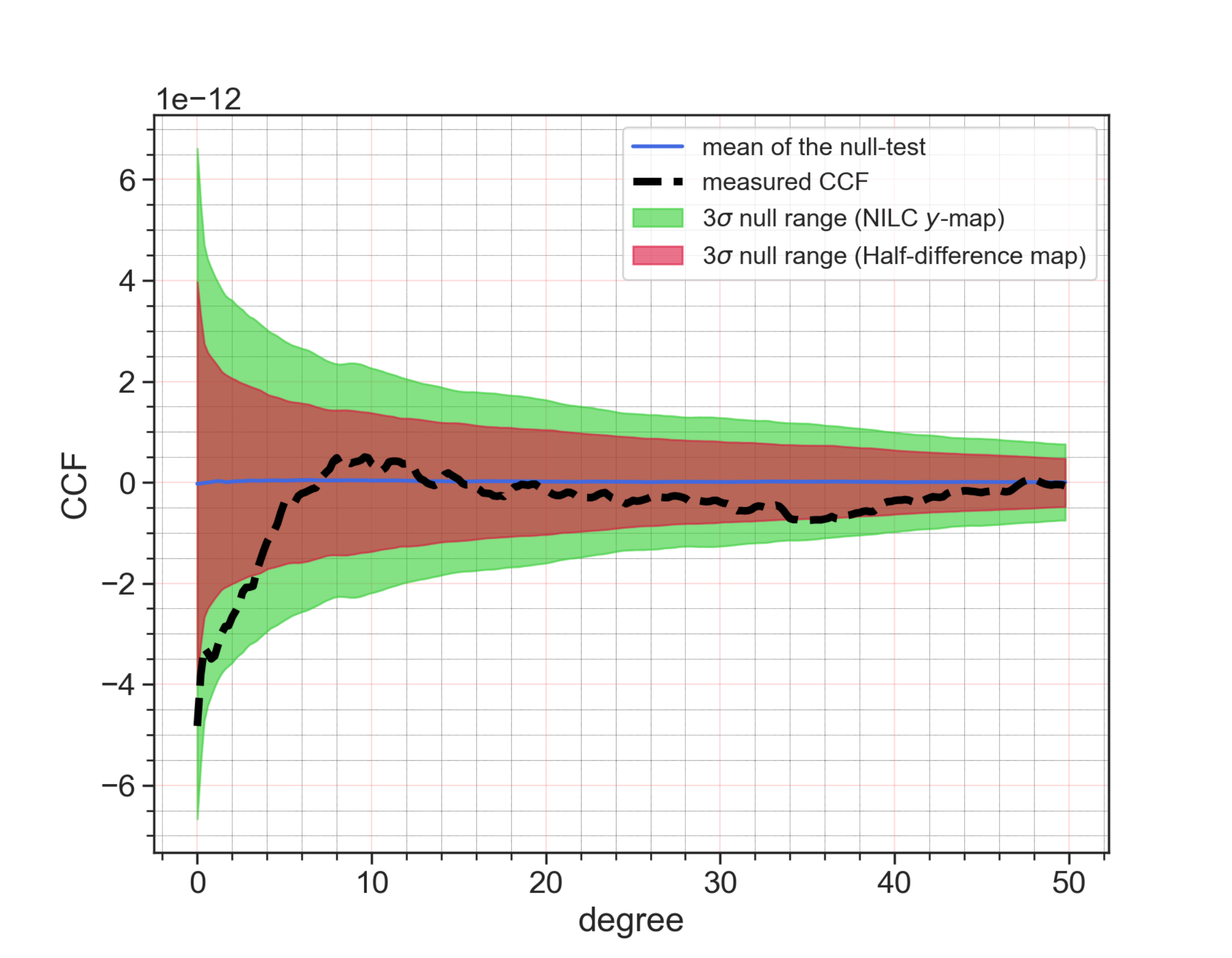}
  \caption{Measured cross-correlation function between masked NILC $y$-map and constructed galaxy density map is shown with black dashed line, as a potential indicator of the existence of WHIM in the filamentary structures around the Virgo Cluster. Green and red coloured regions correspond to 3$\sigma$ range around the mean value (blue straight line) being extracted from null tests by using masked NILC $y$-map and half-difference map, respectively. In other words, these shaded regions correspond to our benchmark to check the significance of the measured signal. It is obvious that the measured signal can't reject the null hypothesis (one-tailed hypothesis test for the expected positive correlation), even when the statistics of half-difference map is exploited, as the signal doesn't exceed the positive side of the 3$\sigma$ range in the benchmark. }
  \label{figure:results_1}
\end{figure}

Since there is an expected positive correlation between the modelled filament $y$-map and constructed galaxy density map 
%(directionally articulated hypothesis) 
as seen in Fig. \ref{figure:expected_CCF}, the significance of the measured CCF between NILC $y$-map and galaxy density map is checked via a one-tailed null-hypothesis test. In Fig. \ref{figure:results_1}, the measured signal appears to be within \textit{non-detection} case, in other words the null hypothesis cannot be rejected. Even when only the instrumental noise statistics (from half-ring maps) is used for the null tests, signal itself is still too weak, which means that noise is too dominant over the related Compton $y$ signal coming from the diffuse gas within these filaments, if any. As is the common standard, statistical significance threshold is set to 3$\sigma$, and corresponding range is shown in Fig. \ref{figure:results_1}. The measured CCF is far too small from exceeding this level throughout the entire range of scales (from 0 to 50 degrees). Moreover, it does not exhibit the expected CCF characteristic as a bump around the scales $\sim$30 - 45 degrees that was specified for analyzing the WHIM signal (Fig. \ref{figure:expected_CCF}). 
%\par
%
%The applied strategy for the null test to check significance level of the measured signal is quite crucial, and that is the main purpose why both null test results are provided in Fig. \ref{figure:results_1} (green and red coloured regions). In our case, a particular statistics of the noise map (half-difference map) and the actual measured field results in a different CCF range, which could effect the main result. The importance of convenient statistics is validated in this way, and also it demonstrates that the masked $y$-map still reserves either the contamination or systematic noise, besides the instrumental noise (right panel in Fig. \ref{figure:Pymaster}). 
In addition, the mean CCF value within the null tests is found to be nearly zero (blue line in Fig. \ref{figure:results_1}), which shows the consistency of the applied method because the expected correlation between random $y$-map and galaxy density map is zero.
\par

Fig. \ref{figure:results_1} validates the use of the masked $y$-map for the purpose of null tests, as it preserves both systematic noise sources and the instrumental noise. Therefore, for further analyses, null-CCF range being extracted by using the actual masked NILC $y$-map itself via Algorithm \ref{nulltest_pseudocode} (green colored region in Fig. \ref{figure:results_1}) will be considered for setting the upper limits on the WHIM physical parameters. 
%, as it automatically comprises a contamination effect and systemics.

%%%%%%%%%%%%%%%%%%%%%%%%%%%%%%%%%%%%%%%%%%%%%%%%%%%%%%%%%%%%%%%\
\subsection{Density upper limits with {\it Planck} data} \label{boosting}

\begin{figure} 
  \includegraphics[width=0.5\textwidth]{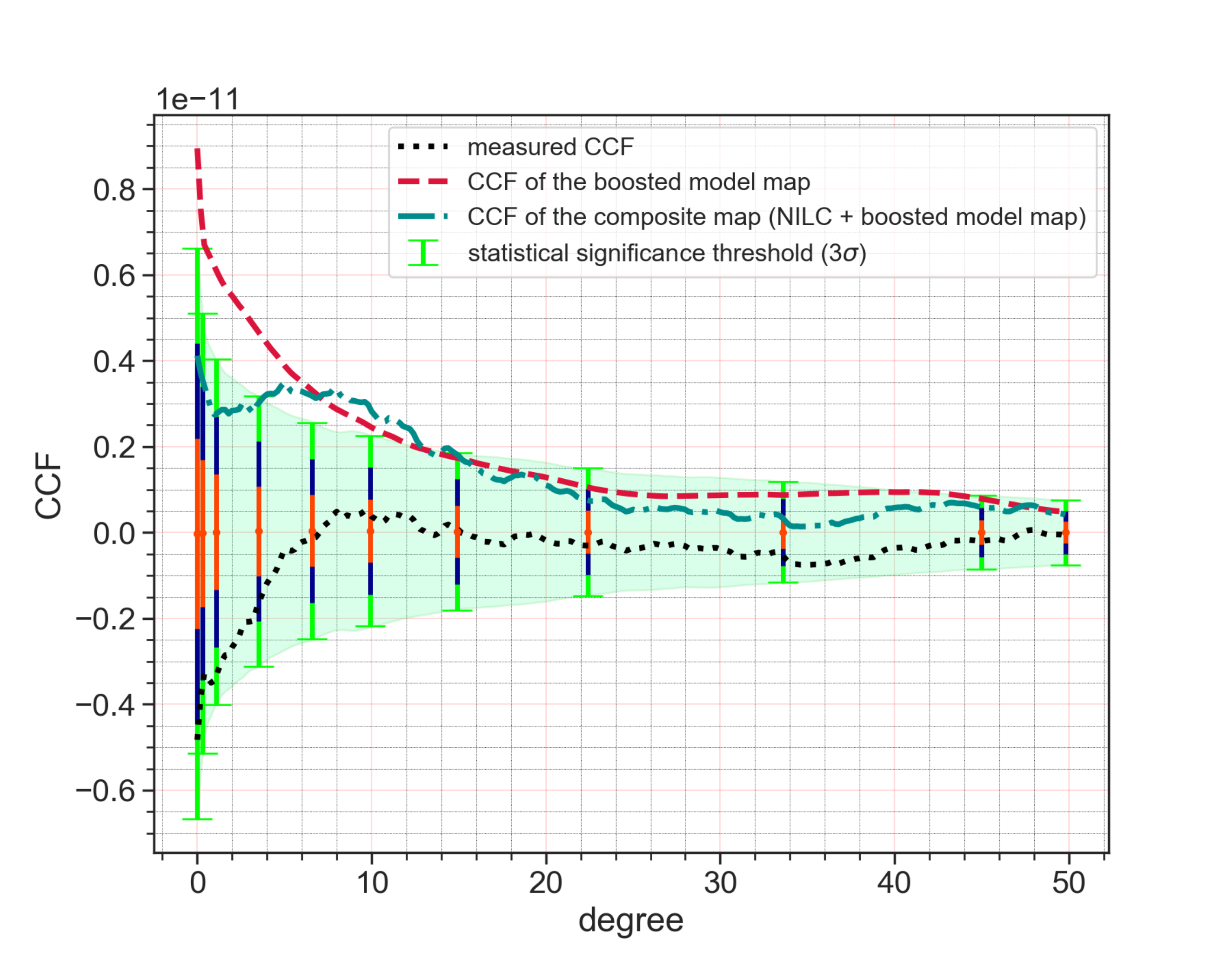}
  \caption{The moment of the null hypothesis rejection, in other words the \textit{detection} case (\cref{boosting}), as the CCF signal of the boosted modelled filament $y$-map (red dashed line) exceeds the 3$\sigma$ level at positive side of the null hypothesis range (green shaded region) within the characteristic scale (gray band in Fig. \ref{figure:expected_CCF}), where the actual measured CCF signal (dotted black line) is far below this level. In fact, measured signal appears to be totally lost as it even couldn't exceed the 1$\sigma$ range (1,2 and 3$\sigma$ intervals for the null hypothesis are shown with different colored bars: orange, blue and green, respectively). Turquoise dash-dotted line corresponds to the CCF signal being extracted from the composite map as a summation of the masked NILC $y$-map and boosted modelled filament $y$-map.}
  \label{figure:boosting}
\end{figure}

We use the non-detection result to constrain density upper limits of the hypothesised WHIM phase within the filamentary structures around the Virgo Cluster. Based on the null-test results (Fig. \ref{figure:results_1}), a positive $3\sigma$ threshold is picked for the \textit{detection} case, i.e., to reject the null hypothesis. 
%At this point, the motivation is that if there was a specific column of WHIM here, then we would measure it against the null test already, but we couldn't currently. 
%
Starting with the reference model filament $y$-map (left panel in Fig. \ref{figure:filament_vs_galaxy_map}), with a relatively low central density value ($n_e(0)$) of the diffuse gas, the strategy is to boost the signal by increasing the axial $n_e(0)$ iteratively, until the calculated CCF between boosted $y$-map and galaxy density map exceeds the $3\sigma$ limit. 

At this point, two options exist regarding the $y$-map that will be used to calculate enhanced signal: either the boosted model $y$-map of the filaments directly, or to employ a composite $y$-map by adding the boosted model $y$-map onto the actual measured NILC $y$-map. The latter approach is used in \citet{brown2017limiting} for limiting magnetic field values in the cosmic web with diffuse radio emission. Besides, the termination criteria from an iteration is also crucial: which interval should be used as a check when the $3\sigma$ limit is exceeded? In \citet{brown2017limiting}, there is no constraint regarding this criteria, the iteration is terminated when the boosted CCF signal exceeds $3\sigma$ at \textit{any} degree. In \citet{Tanimura2019a}, likelihood of the data exceeding a null hypothesis is calculated by comparing the $\chi^2$ of the measured data to the $\chi^2$ of the probability distribution function of 1000 null samples. 

\par
In our case, the boosted modelled filament $y$-map can directly be used as a reference. In other words, there is no need to contaminate a reference $y$-map with the noise-dominant $y$-map (i.e., the masked NILC $y$-map, Fig. \ref{figure:results_1}). Nevertheless, derivation of the density threshold numbers are obtained using both techniques, and we note some important points regarding both. 
Firstly, boosted signal from the composite $y$-map may can bear the trace of correlated noise (instrumental or other systematic), as can be seen in Fig. \ref{figure:boosting} (turquoise dash-dotted line). This can complicate the interpretation of the density limit values.  
%Actually, deviation of the extracted density upper limit value is calculated by using both of the techniques and the result is provided below. 
Secondly, specific characteristic of the expected CCF (the bump between roughly 30 and 45 degrees, Fig. \ref{figure:expected_CCF}) is exploited to eliminate noise contamination effect, especially when using the  composite map (Fig. \ref{figure:boosting}). 
%and also to be able to provide preserved causality as it reflects the topological nature of the analysed filaments around the Virgo Cluster specially. 

For each constructed bin with a width of 0.2 degree throughout the angular range (0 to 50 degrees), null samples have a Gaussian distribution centered on zero, so p-value calculation according to chi-square test for null-test rejection is equal to the direct $3\sigma$ thresholding. As a consequence, the optimized boosting algorithm is as follows: the modelled filament $y$-map is initialized with relatively low central density and then it is iteratively increased until the calculated CCF exceeds the 3$\sigma$-level of null-range between $30^{\circ}$ and $45^{\circ}$. The CCF is calculated between galaxy density map and boosted modelled $y$-map directly, not combined with NILC $y$-map at this point, however, results are also obtained with the composite $y$-map usage for robustness check. Within the iterative loop, to obtain the density value corresponding to the limit for null hypothesis rejection, adapted version of the Newton-Raphson method (first-order in the class of Householder's methods which are numerical algorithms for root-finding) is used with a tolerance of $\Delta n_e(0) \leq 3\, \text{m}^{-3}$. 
\par

Fig. \ref{figure:boosting} (red-dashed line) just captures the moment where the CCF between boosted modelled filament $y$-map and galaxy density map exceeds $3\sigma$ threshold at the positive side (\textit{detection} case) with a tolerance of $\Delta n_e(0) \leq 3\, \text{m}^{-3}$, where the obtained central and mean (transverse) electron densities are $n_e(0) \approx 3 \times 10^{-4} \, \text{cm}^{-3}$ and $\langle n_e \rangle \approx 1.25 \times 10^{-4}\, \text{cm}^{-3}$, respectively. Furthermore, this bump-based iterative boosting algorithm is already providing the similar null hypothesis rejection ability when an overall chi-square test is applied (with calculation of p-value according to $\chi^2$ PDF of null CCF profiles), as the boosted CCF exceeds $3\sigma$ in wide range of scales too, and at least above $2\sigma$ level for the entire range (excluding the very small angles close to the resolution limit of the model map, where the autocorrelation signal blows up).

\par
In the \textit{detection} case mentioned above, corresponding CCF profile extracted from the composite $y$-map (masked NILC $y$-map $+$ boosted modelled $y$-map) is also shown in Fig. \ref{figure:boosting} (turquoise dash-dotted line). It's noticeably affected by random correlations within the iterative process, as its shape isn't smoothly evolving while boosting (i.e. not conserving its noise-dominant profile shape), and this might cause a wrong interpretation when the profile exceeds $3\sigma$ level at any degree. This behaviour obviously shows the critical importance of eliminating noise characteristics and instrument dependencies. Nevertheless, the density upper limit value applying the same algorithm is obtained by using this boosting technique (with the composite $y$-map rather than only the modelled $y$-map itself). As the intrinsic noise results in a confusion (via triggered random damping), to be able to exceed $3\sigma$ level at the bump region the composite map requires more boosting than the default one. However, the difference is not considerable: the found upper limit value is  roughly 10$\%$ larger than the default one ($\langle n_e \rangle \approx 1.40 \times 10^{-4}\, \text{cm}^{-3}$). Hence, the main outcome is that the obtained density upper limit does not strongly depend on the chosen boosting technique, which provides a degree of robustness for our obtained limits. Because of this finding, for further analysis in \cref{error_estimation}, noise-free $y$-map (boosted modelled filament $y$-map) is used for null hypothesis rejection.

%%%%%%%%%%%%%%%%%%%%%%%%%%%%%%%%%%%%%%%%%%%%%%%%%%%%%%%%%%%%%%%\
\subsection{Error Estimation $\&$ Robustness Check} \label{error_estimation}

\begin{figure} 
  \includegraphics[width=0.5\textwidth]{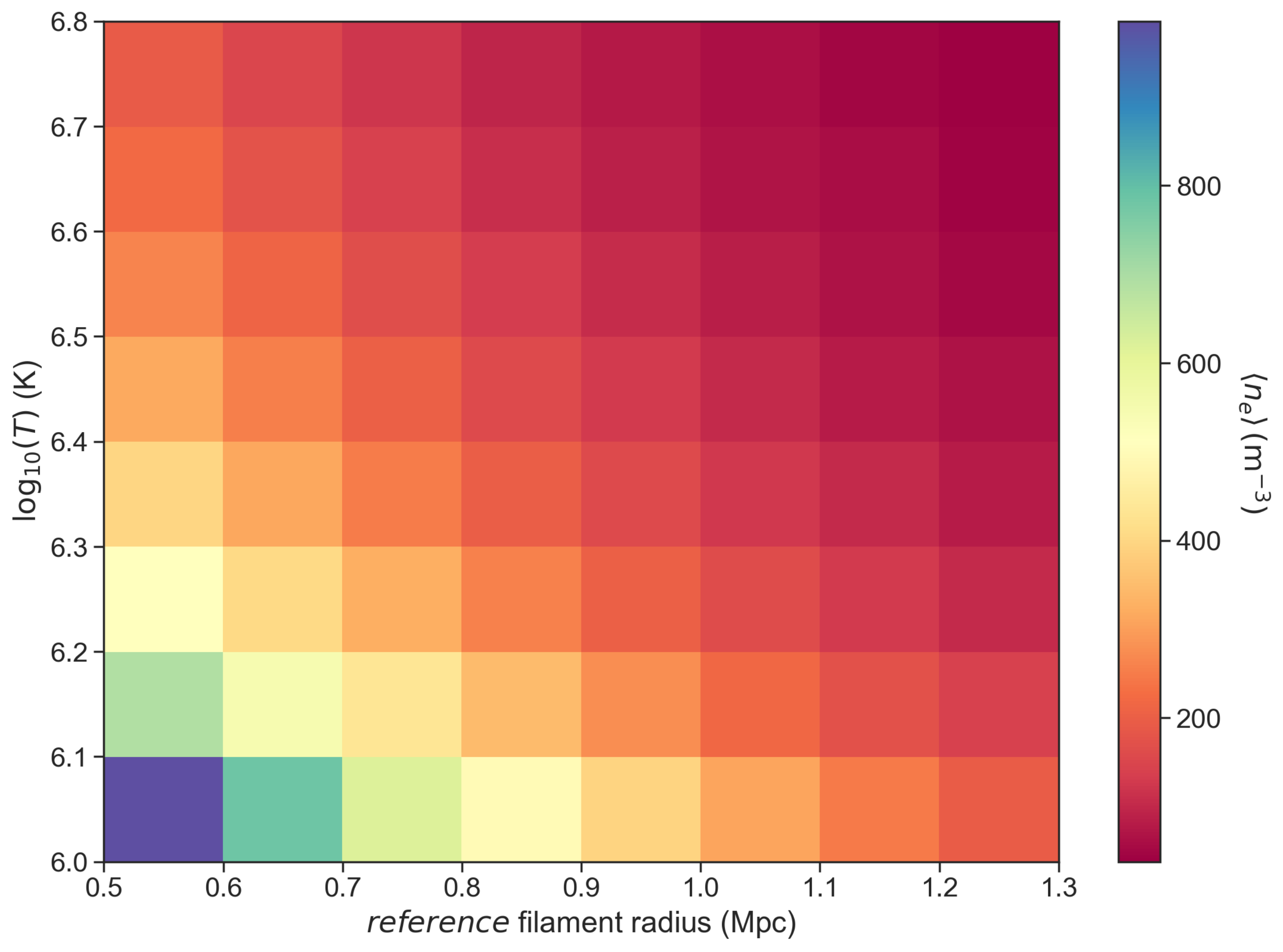}
  \caption{Constructed mesh grid being colored by the obtained density upper limits $\langle n_e \rangle$ in $\text{m}^{-3}$. The axes are filament temperature in logarithmic scale (isothermal assumption) versus reference (NGC 5353/4) filament radius in Mpc., which are used while modelling the filaments $y$-map (left panel in Fig. \ref{figure:filament_vs_galaxy_map}).}
  \label{figure:meshgrid}
\end{figure}

While computing power spectrum of a given sky patch, the variance on the power value, especially at lower multipoles, must include the effect of cosmic/sampling variance. However, in our case, full-sky $C_l$ estimator (\textcolor{gray}{NaMaster}) is used for obtaining randomly-realized maps that have similar statistics on the same specific sky patch. 
%Besides, the correlation analysis takes place on our local universe, centered around the Virgo cluster with a highly masked region. 
Thus, cosmic variance is already included in the error estimates. In fact, as it can be seen from the left panel in Fig. \ref{figure:Pymaster}, generated power spectra of the simulated maps are already scattering around the actual measured spectrum of the masked NILC $y$-map, which thus provides the requisite tolerance of the measured reference power spectrum from cosmic/sampling variance.

\par
Therefore, the dominant factors regarding error estimation all concern with the modelling of the WHIM within the filamentary structures (\cref{filament_ymap}): chosen density profile, filament radius, and assumed temperature isothermality. There is no universally adopted density profile for the WHIM as it depends on the baryonic physics adopted in the simulations. 
%General estimation has to be done at this point, and relatively shallow one which is characteristically seen in many simulations regardless of the used baryonic physics to simulate filaments (\cref{filament_ymap_3}) is picked. 
$\beta$-model fit is motivated for obtaining a shallow density with a smooth inner region and a constant power-law slope outwards for the diffuse gas. 
%Thus, just tried to mimic a typical profile being extracted from hydrodynamical simulations (\cref{filament_ymap_3}), and fitted profile parameters aren't altered on the purpose of error estimation for this work. 
As the essential drivers for our derivation of density upper limit are filament radius and temperature, we construct a mesh-grid for the density values based on these two parameters, which can be seen in Fig. \ref{figure:meshgrid}. 
Range of these parameter values is selected according to the noticed scatter in the scaling relations which are mentioned as a part of filament $y$-map reconstruction (\cref{filament_ymap}). 

For Fig. \ref{figure:meshgrid}, the boosting algorithm (\cref{boosting}) is applied for a given parameter combination, and the resulted grid is colored by the obtained density upper limits ($\langle n_e \rangle$). As is expected, density upper limit increases while the temperature and radius decreases, to be able to produce the same level of correlation signal. The range of obtained $\langle n_e \rangle$ limits are between $\approx3 \times 10^{-5} \text{cm}^{-3}$ (for $T=10^{6.8}\, \text{K}$ $\&$ $r=1.3 \, \text{Mpc.}$) and $\approx9 \times 10^{-4} \text{cm}^{-3}$ (for $T=10^{6}\, \text{K}$ $\&$ $r=0.5\, \text{Mpc.}$). 
%Ideally, for a given parameter space, one should also has a weight table to be able to calculate weighted mean and variance for the analysis, but it's not currently possible at this point. 
Our approach is providing a conservative upper limit, to enable one to state that it is unlikely to exceed this mean density value for a diffuse WHIM gas within these filaments. To calculate a representative mean, rank-based $3\sigma$ clipping (trim 5$\%$ of the flattened $\&$ sorted matrix which represents the high end, left-bottom triangle of the meshgrid in Fig. \ref{figure:meshgrid}) is applied onto the sorted upper limit values. The calculated mean density limit is $\langle n_e \rangle \leq 4 \times 10^{-4}$ cm$^{-3}$, roughly 3 times the value found in \cref{boosting}.

\par
The robustness of the density upper limit range is also checked in terms of the \plk Collaboration pipeline that was used for Compton $y$-map generation, and also for the smoothing kernel size (FWHM) in galaxy density map construction. For the first, MILCA $y$-map \citep{aghanim2016planck-tSZ} is used rather than NILC $y$-map within each step of the overall algorithm consist of masking, simulation of randomly realized maps, and boosting until null-rejection. Also, galaxy density map is built by using different size of kernels with $20^{\prime} \,\&\, 40^{\prime}$ of FWHM. The resulting differences on the obtained mean density upper limits are minor, and therefore, their effect can be considered as negligible because the amount of divergence is smaller than inter-square variations between adjacent cells visualized in Fig. \ref{figure:meshgrid}.

%%%%%%%%%%%%%%%%%%%%%%%%%%%%%%%%%%%%%%%%%%%%%%%%%%%%%%%%%%%%%%%%\
\section{Discussion and Conclusions} \label{discussion_and_conclusions}
For the modelling of the WHIM within the filamentary structures around the Virgo Cluster, certain simplifications are made on the characteristic properties whose effects are ignored for deriving the density upper limit values. Below is a list of unaccounted factors which might have impact on the small scale correlation strength: 
\begin{itemize}
	\item[$-$] There is a weak correlation between gas mass and temperature in filaments, as described by the $M$ - $T$ scaling relation in \citet{gheller2015properties}, but temperature is considered as independent from the mass of the filaments in our case.
	\item[$-$] Lengthwise changes in density profile along the spine axis are ignored, and homogeneity is assumed throughout our analysis (i.e., no gas clumping). 
	\item[$-$] Fixed density profile is adapted for each filament, and also the same temperature value is shared, to exploit self-similar behaviour for the  filaments, even though Virgo is a dynamically active galaxy cluster. 
	\item[$-$] Filaments might have irregular geometries, depending on the environment or their evolutionary stages \citep{gheller2015properties}. In our case, they are implemented as smooth cylinders with a fixed  diameter in each individual cases.
	\item[$-$] Some amount of \textit{bound} gas around over-dense objects, such as the galaxy halos within the filaments, can also be expected \citep{gheller2019survey}. This can contribute to an additional correlation signal in the measured CCF. However, this contribution is also well within the noise limit, as already can be seen from Fig. \ref{figure:results_1}. Besides, the derived upper limits in \cref{error_estimation} can be considered conservative in the context of this issue, since it does not account for the extra contribution of bound gases towards the correlation signal.
\end{itemize}\par

%%Discussion on Virgo-centric distance
%%
We have assumed a flat $\Lambda$CDM cosmology, given by the Planck 2013 results \citep{ade2014planck}, to convert the redshift listed in the optical galaxy catalog to distances and to construct the 3D filament model. This choice leads to a mean Virgo-centric distance of 18.3 Mpc. Since there is still some debate on the exact value of Virgo from the Milky Way reference frame (a generally adopted value is 16.5 Mpc, see \citet{Mei2007}, we should consider if an error in the assumed Virgo-centric distance could affect the results from the cross-correlation analysis. Firstly, we note that even if there is an overall shift in the Virgo-centric distance, the sky coordinates remain the same, and hence our 2D CCF results based on the $y$-map are unaffected. This also includes the null-test results for error analysis. However, when converting the results into density limits, there can be a small change in the values due to a resulting change in the volume. We consider this effect to be fairly small, given the roughly 10\% difference in the absolute distance scale of Virgo that can be expected, and also due to the effect of projection. As such, any changes in the density upper-limit values should be well below the accuracy permitted by the data.

There might be another potential effect from the pipeline used in the identification of galaxies in cosmic filaments around the Virgo Cluster. In \citet{Kim2016}, 2$\sigma$ clipping is applied to the galaxies around the fitted spine axis to extract filament-hosted galaxy members. This number of galaxies is taken as a reference parameter in the scaling relations to calculate filament radius rates in \cref{filament_ymap}. Nevertheless, as the galaxy number ratio of filaments is critical at this point rather than the actual member numbers, the clipping procedure does not play a crucial role for the mass proxy via scaling relations in this study. 
Besides, \citet{gheller2015properties} used a criterion based on the mass density threshold in simulation cells for the identification of filaments, and in their Appendix A it's shown that the main trends and dispersions obtained for different threshold values are similar, and also scaling relations are consistent.

%Besides, \citet{gheller2015properties} used the following criteria for an identification of filaments in simulations
%\begin{equation}
%\frac{\rho_\mathrm{BM}}{\rho_0} \ge a_\mathrm{fil}
%\end{equation} where $\rho_{BM}$ is baryon mass density, $\rho_0$ is critical density at present time, and $a_\mathrm{fil}$ is a mass density threshold for the cells to be assigned as a part of filament structures. Herein, $a_\mathrm{fil}$ is the most important parameter which could influence retrieved statistical properties of filaments, e.g. transverse size distribution, and so in this way the extracted scaling relations too. Most especially, the effect on properties like average density profile is found to be neglectable. Besides, in Appendix A of \citet{gheller2015properties}, it is shown that the main trends and dispersions being obtained for different thresholds are similar, and also scaling relations are consistent.

In this study, the main result is a representative range of the density upper-limit values, obtained via varying the cross-correlation signal amplitude, by means of using models of Compton $y$-parameter, filament radius, and temperature.  
%Then, the conservative value is obtained based on the extracted values from these factors. The above-mentioned points are assumed to be negligible for a first-order estimation of the density upper limits. It is also worth noting that the range of density limits that we derive from the current \plk data are right at the boundary of the WHIM density range as recently simulated by \citet{Martizzi2019}. 
The mean density upper limit at 3$\sigma$ detection level found in \cref{error_estimation} is totally consistent with the WHIM parameter space extracted from simulations such as Fig. 4 in \citet{martizzi2019baryons} and Fig. 17 in \citet{gheller2015properties}. The range being shown in Fig. \ref{figure:meshgrid} is particularly well-matched with the expected density range for filamentary structures \citep{martizzi2019baryons}. This raises the exciting possibility that the next generation of space-based multi-wavelength CMB experiments, like the proposed LiteBIRD mission \citep{LiteBIRD} which would be able to measure the Compton-$y$ signals from large angular scales and with better precision, would be able to detect the CCF signal that we have discussed from the Virgo filaments, and possibly also from many other nearby cluster systems. An overall reduction by a factor $2-3$ of the combined instrumental noise and systematic noise residuals (astrophysical foregrounds) should be sufficient for that goal, which is within the design criteria of missions like LiteBIRD. Even if there is a non-detection result as in this paper, the constraining power of the CMB data from these future missions will enable the relaxation of some of the simplifying assumptions mentioned above, to obtain consistent results with the hydrodynamic simulations, and this would contribute to the critical evaluation of  WHIM-formation theories.\\ \\

%%%%%%%%%%%%%%%%%%%%%%%%%%%%%%%%%%%%%%%%%%
\begin{acknowledgements}
      We thank Franco Vazza for many helpful suggestions regarding baryonic properties of cosmic filaments. We thank Jens Erler, Mathieu Remazeilles, Frank Bertoldi for helpful discussions.
\end{acknowledgements}

%%%%%%%%%%%%%%%%%%%%%%%%%%%%%%%%%%%%%%%%
% REFERENCES
\bibliographystyle{aa} % style aa.bst
\bibliography{whimpapers} % your references Yourfile.bib

%%%%%%%%%%%%%%%%%%%%%%%%%%%%%%%%%%%%%%%%%%%%%%%%%%%%%%%%%%%%%%
% APPENDIX
\begin{appendix}

\section{Correlation function from the angular power spectrum}
\label{app:corr_function}

By following the HEALPix conventions, a bandlimited function $f$ defined over the spherical surface grid can be expanded in spherical harmonics, $\textit{Y}_{\textit{lm}}$, as
\begin{equation}
f(n) = \sum\limits_{l=0}^{l_{\textit{max}}} \sum\limits_{m} a_{lm}Y_{lm}(n)
\end{equation}
where $n$ denotes a unit vector pointing at polar angle $\theta \in [0,\pi] $ and azimuth $\phi \in [0,2\pi)$. Pixelated $f(n)$ corresponds to sampling $f$ at $N_{pix}$ locations $n_p$, $p \in [0, N_{pix} - 1]$. Then, this sampled function can be used to estimate $a_{lm}$,
\begin{equation} \label{a_lm}
\hat{a}_{lm} = \frac{4\pi}{N_{pix}} \sum \limits_{p=0}^{N_{pix}-1} Y^*_{lm}(n_p)f(n_p)
\end{equation}
where the superscript star denotes complex conjugation, and an equal weighting scheme is assumed for each pixel. Then, these coefficients which correspond to Fourier amplitudes can be used to compute estimates of the angular power spectrum $\hat{C}_l$ as
\begin{equation} \label{Cl}
\hat{C}_{l} = \langle a_{lm} a^*_{lm}\rangle = \frac{1}{2l+1} \sum \limits_{m} |\hat{a}_{lm}|^2
\end{equation}
Using this definition we can rewrite the correlation function in spherical harmonics
\begin{equation} \label{correlation_func}
\begin{split}
\xi\left(\Vec{x}\right) \equiv \langle f\left(n\right)f\left(n'\right)\rangle = \sum \limits_{l}C_l \sum \limits_{m} Y_{lm}(n)Y^*_{lm}(n')\\
= \sum \limits_{l} \frac{2l+1}{4\pi} C_l P_{l}\left(cos(\theta)\right)
\end{split}
\end{equation}
where $\theta$ is the angle between the two directions and $P_l$ is the Legendre polynomial which corresponds to sum over $m$ indices in Eq. \ref{correlation_func} (see Appendix A.4 in \citet{gorski1999healpix}). In \textcolor{gray}{HEALPY} module, \textcolor{gray}{anafast} computes the power spectrum (Eq. \ref{a_lm} $\&$ \ref{Cl}) and then \textcolor{gray}{bl2beam} computes the angular profile from its transfer function $\hat{C}_l$ (Eq. \ref{correlation_func}). Besides, one can use \textcolor{gray}{synfast} to generate a random realisation of $f(n_p)$ from its input power spectrum $C_l$, which is also used within the null tests in cross-correlation analysis (\cref{null_tests}). \par
\end{appendix}
\end{document}